\documentclass[gmd, hvmath]{copernicus}
\usepackage{xcolor}
\usepackage{widetext}
\begin{document}\sloppy

\title{Improving predictive power of physically based rainfall-induced shallow
  landslide models: a~probabilistic approach}
\author[1]{S.~Raia}
\author[1]{M.~Alvioli}
\author[1]{M.~Rossi}
\author[2]{R.~L.~Baum}
\author[2]{J.~W.~Godt}
\author[1]{F.~Guzzetti}

\affil[1]{CNR IRPI, via Madonna Alta 126, 06128 Perugia, Italy}
\affil[2]{US Geological Survey, P.O.~Box 25046, Mail Stop 966, Denver, CO 80225-0046, USA}

\runningtitle{Improving landslide modeling}
\runningauthor{S.~Raia et~al.}

\correspondence{M.~Alvioli (alvioli@pg.infn.it)}

\received{7~January 2013 }
\accepted{12~February 2013}
\published{}

\firstpage{1}

\maketitle

\begin{abstract}
      Distributed models to forecast the spatial and temporal occurrence of
      rainfall-induced shallow landslides are based on deterministic laws.
      These models
      extend spatially the static stability models adopted in geotechnical
      engineering, and adopt an infinite-slope geometry to balance the
      resisting and the driving forces acting on the sliding mass. An
      infiltration model is used to determine how rainfall changes
      pore-water conditions, modulating the local stability/instability
      conditions. A~problem with the 
      operation of the existing 
      models lays in 
      the difficulty in
      obtaining accurate values for the several variables that describe the
      material properties of the slopes. The problem is particularly severe
      when the models are applied over large areas, for which sufficient
      information on the geotechnical and hydrological conditions of the
      slopes is not generally available. To help solve the problem, we
      propose a~probabilistic Monte Carlo approach to the distributed
      modeling of rainfall-induced shallow landslides. For the purpose, we
      have modified the Transient Rainfall Infiltration and Grid-Based
      Regional Slope-Stability Analysis (TRIGRS) code. The new code
      (TRIGRS-P) adopts a~probabilistic approach to compute, on a~cell-by-cell
      basis, transient pore-pressure changes and related changes in the
      factor of safety due to rainfall infiltration. Infiltration is modeled
      using analytical solutions of partial differential equations
      describing one-dimensional vertical flow in isotropic, homogeneous
      materials. Both saturated and unsaturated soil conditions can be
      considered. TRIGRS-P copes with the natural variability inherent to
      the mechanical and hydrological properties of the slope materials by
      allowing values of the TRIGRS model input parameters to be sampled
      randomly from a~given probability distribution. The range of variation
      and the mean value of the parameters can be determined by the usual
      methods used for preparing the TRIGRS input parameters.  The outputs
      of several model runs obtained varying the input parameters are
      analyzed statistically, and compared to the original (deterministic)
      model output. The comparison suggests an improvement of the predictive
      power of the model of about 10\,\unit{\%} and 16\,\unit{\%} in two
      small test areas i.e., i.e., the Frontignano (Italy) and the Mukilteo (USA)
      areas. We discuss the computational requirements of
      TRIGRS-P to determine the potential use of the numerical model to
      forecast the spatial and temporal occurrence of rainfall-induced
      shallow landslides in very large areas, extending for several hundreds
      or thousands of square kilometers. Parallel execution of the code
      using a~simple process distribution and the Message Passing Interface
      (MPI) on multi-processor machines was successful, opening the possibly
      of testing the use of TRIGRS-P for the operational forecasting of
      rainfall-induced shallow landslides over large regions.
\end{abstract}

\introduction\label{sec:introduction}

      Rainfall is a~primary trigger of landslides, and rainfall-induced
      landslides are common in many physiographical environments,
      e.g. \cite{brabb1989}. Prediction of the location and time of
      occurrence of shallow rainfall-induced landslides remains
      a~difficult task, which can be accomplished adopting empirical
\citep{crosta1998,sirangelo2003,aleotti2004,guzzetti2007,guzzetti2008},
statistical (\citeauthor{soeters1996}, \citeyear{soeters1996};
\citeauthor{guzzetti1999}, \citeyear{guzzetti1999}, \citeyear{guzzetti2005}, \citeyear{guzzetti2006a};
Committee on the Review of the National Landslide Hazards Mitigation Strategy,
2004), or process based \citep{montgomery1994,terlien1998,baum2002,baum2008,baum2010,crosta2003,simoni2007,godt2008,vieira2010} approaches, or a~combination of
them \citep{gorsevski2006,frattini2009}.
Inspection of the literature, reveals that process based
(deterministic, physically based) models are preferred to forecast the
spatial and the temporal occurrence of shallow landslides triggered by
individual rainfall events in a~given area. Process
based models rely upon the understanding of the physical laws controlling
slope instability. 
Due to lack of information and the poor understanding of
the physical laws controlling landslide initiation, only simplified,
conceptual models are possible, currently. 
These models extend spatially the simplified stability models widely adopted
in geotechnical engineering e.g., \cite{taylor1948,wu1995,wyllie2004}, 
and calculate the stability / instability 
of a~slope using 
parameters such as normal stress, angle of internal friction, cohesion,
pore-water pressure, root strength, seismic acceleration, external weights.
Computation results in the factor of safety, an index expressing the ratio
between the local resisting ($R$) and driving ($S$) forces, $F_{\mathrm{S}}= R / S$.
Values of the index smaller than 1, corresponding to $R < S$, denotes instability,
on a~cell-by-cell basis, according to the adopted model. To calculate the 
resisting and the driving forces, the geometry of the sliding mass must be defined,
including the geometry of the topographic surface and the location of the 
slip surface. Most commonly, an infinite-slope approximation is adopted
\citep{taylor1948,wu1995}. 
This is also the approach adopted by the US Geological Survey (USGS) 
Transient Rainfall Induced and Grid-Based Regional Slope-Stability Model (TRIGRS) model
\citep{baum2002, baum2008}, within each user-defined cell. Within the infinite-slope
approximation, in each cell the slip surface is assumed to be of infinite extent, planar,
at a~fixed depth, and parallel  to the topographic surface. Forces acting on the sides of
the sliding mass are neglected.
\begin{figure}[!htp]%
  \includegraphics[width=60mm]{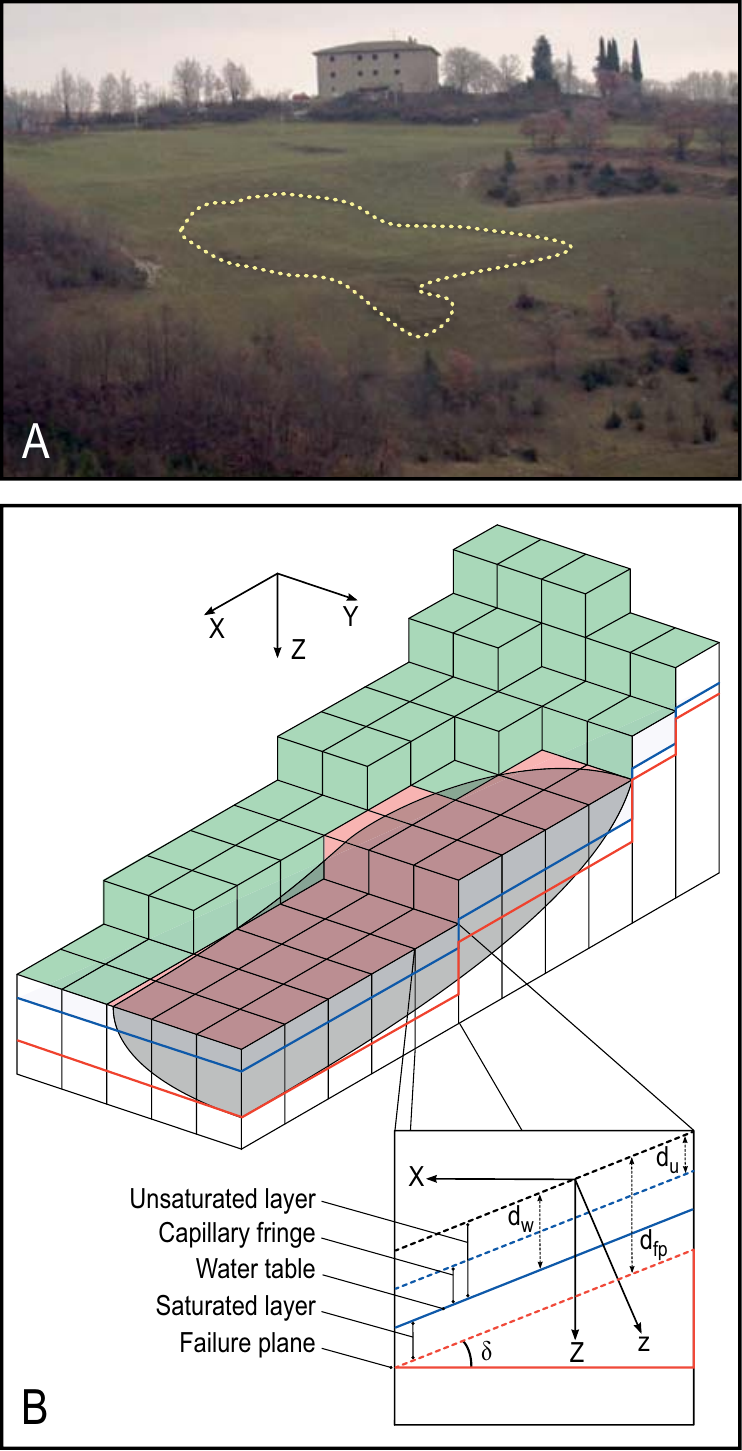}
  \caption{%
    \textbf{(A)} Example of a~rainfall-induced shallow landslide of the soil slide
    type in the Collazzone area, Umbria, Italy (Figure~\ref{fig07}). \textbf{(B)} Schematic
    representation of the slope-infinite model showing the coordinate system and
    variables used in the deterministic and probabilistic models. See Table~\ref{notation}
    for the symbols description.}
  \label{fig01}%
\end{figure}%
      Modeling of shallow landslides (Figure~\ref{fig01}a) triggered by
      rainfall adopting the infinite-slope approach requires time-invariant
      and time-dependent information.  Time-invariant information includes
      the mechanical and hydrological properties of the slope material
      (e.g. unit weight $\gamma$, cohesion $c$, angle of internal friction
      $\phi$, water content $\theta$, saturated hydraulic conductivity
      $K_{\mathrm{s}}$), and the geometrical characteristics of the sliding
      mass (e.g., gradient of the slope and the sliding plane $\delta $,
      depth to the sliding plane $d_{\text{fp}}$).  The fact that these
      parameters are constant in time is an assumption of the model.
      Time-dependent information consists of the pressure head $\psi$,
      i.e. the pressure exerted by water on the sliding mass, a~function of
      the depth, $d_{\mathrm{w}}$ of water in the terrain
      \citep{freeze1979}. Determining the pressure head, and its spatial and
      temporal variations, requires understanding how rainfall infiltrates
      and water moves into the ground. This is described by the Richards
      equation \citep{richards1931}. This non-linear partial differential
      equation does not have a~closed-form analytical solution, and
      approximate solutions are used for saturated
      \citep[e.g.,][]{iverson2000} and unsaturated
      \citep[e.g,.][]{srivastava1991,savage2003, savage2004} conditions.

      The numerical implementation of one such model has been accomplished
      by \citet{baum2002} in TRIGRS. The program calculates the stability
      conditions of individual grid cells in a~given area, and models
      infiltration adopting the approach proposed by \citet{iverson2000},
      for one-dimensional vertical flow in isotropic, homogeneous materials,
      and for saturated conditions. In the code, the forces acting on each
      indi\-vi\-dual grid cell are balanced in the centre of mass of each
      cell, and all interactions with the neighboring grid cells are
      neglected.

      In a~second release of TRIGRS, \citet{baum2008} have extended the code
      to include unsaturated soil conditions, including the presence of
      a~capillary fringe above the water table. TRIGRS can be used for
      modeling and forecasting the timing and spatial distribution of
      shallow, rainfall-induced landslides in a~given area \citep{baum2002,
      baum2008, baum2010}. A~problem when using TRIGRS, and similar computer
      codes (e.g.  Shalstab, \citeauthor{montgomery1994},
      \citeyear{montgomery1994}; GEOtop-SF, \citeauthor{simoni2007},
      \citeyear{simoni2007}), for the modeling of shallow rainfall-induced
      landslides over large areas resides in the difficulty (or operational
      impossibility) of obtaining sufficient, spatially distributed
      information on the mechanical and hydrological properties of the
      \begin{figure}[!htp]%
        \includegraphics[width=80mm]{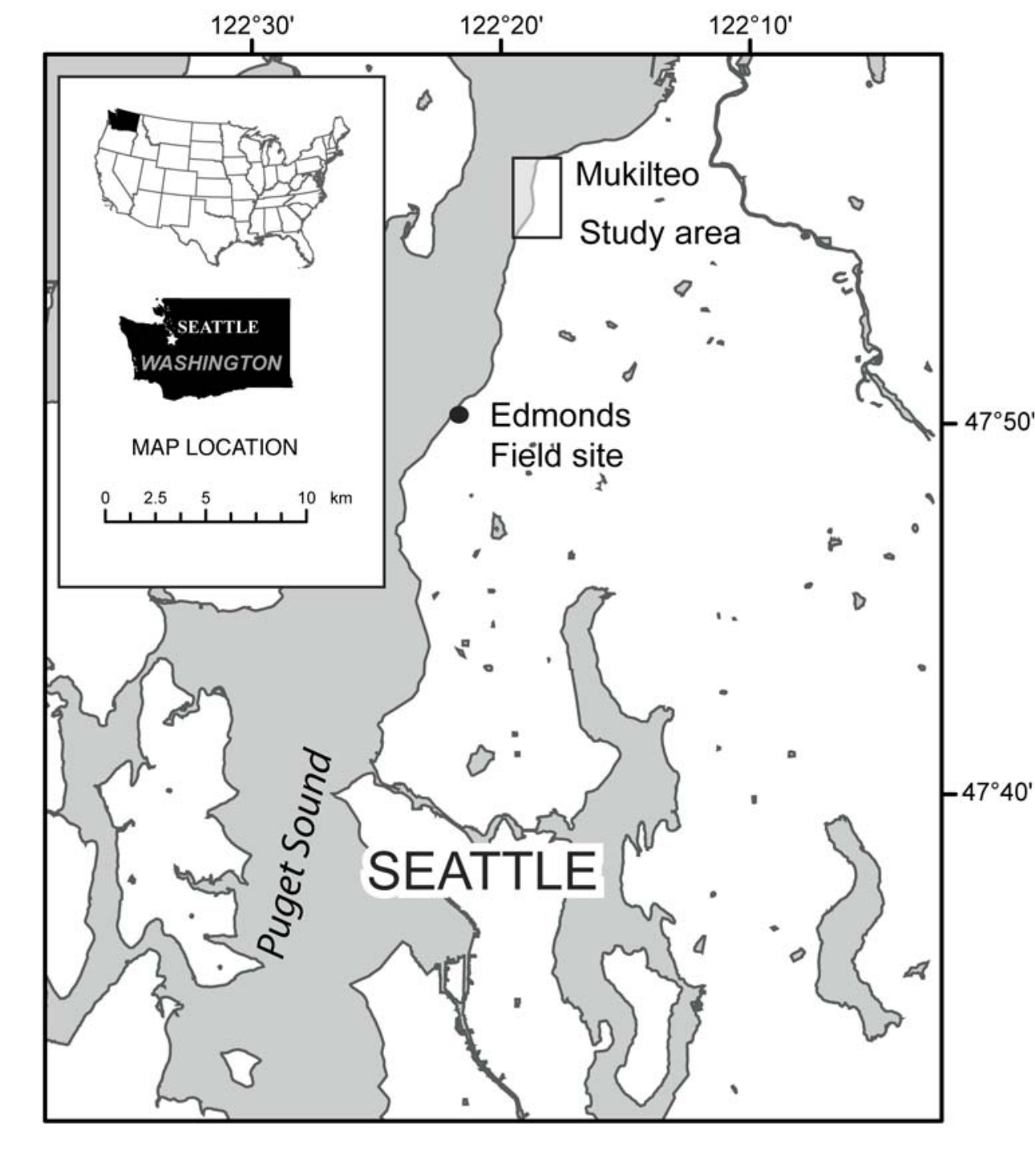}
        \caption{%
          The location of the Mukilteo study area, near Seattle, WA, USA.}
        \label{fig02}%
      \end{figure}%
      terrain. Adoption of a particular value to describe the
      mechanical (unit weight $\gamma_{\mathrm{s}}$, cohesion $c$, angle of
      internal friction $\phi$) and the hydrological (water content
      $\theta$, saturated hydraulic conductivity $K_{\mathrm{s}}$)
      properties of the terrain may result in unrealistic or inappropriate
      representations of the stability conditions of individual or multiple
      grid cells.

      In this work, we propose a~probabilistic, Monte Carlo approach in an
      attempt to overcome the problem of poor knowledge of terrain
      characteristics over large study areas. We obtain the input values for
      the parameters for the individual runs of TRIGRS using probability distributions. 
      Multiple simulations are performed with different sets of randomly chosen input
      parameters, and we obtain multiple sets of model outputs.
      We denote the newly developed code TRIGRS-Probabilistic, or TRIGRS-P. The
      different outputs are then analysed jointly to infer local stability
      or instability conditions as a~function of the random variability of
      the input parameters, and the statistical significance of the multiple
      outputs is determined.
      Examples of similar probabilistic approaches to model the stability/instability conditions 
      of slopes exists in the literature (e.g., \citet{hammond1992,pack1998,haneberg2004}). 
      The various models adopt different physically-based models, which are not equivalent. 
      We maintain that the probabilistic approach of the modified version of 
      TRIGRS is relevant, because it considers most  of the 
      aspects relevant to slope stability analysis, and it is capable of reproducing 
      empirical properties of rainfall-induced shallow landslides, 
      including the rainfall intensity-duration conditions that generate the slope instabilities, 
      and the statistics of the size of the unstable areas, as recently shown by \citet{alvioli2014}. 

      The paper is organized as follows. First, we summarize the model
      adopted in the software code TRIGRS, version 2.0 \citep{baum2008}, and
      we introduce our probabilistic extension (Sect.~\ref{sec:models})
      implemented in the new code TRIGRS-P. Next, we present a~comparison of
      the performance of the original and the probabilistic simulations for
      two study areas: Mukilteo, USA, and Frontignano, Italy
      (Sect.~\ref{sec:model_res}). Results are discussed in
      Sect.~\ref{discussion}, which focuses on the analysis of the
      performance of the geographical prediction of the shallow landslides,
      and on the potential application of the new probabilistic code for
      modeling shallow rainfall-induced landslides over large areas ($>$\,100\,\unit{km^2}).

\section{Overview of the model}\label{sec:models}

      Both TRIGRS and TRIGRS-P frameworks are pixel-based and adopt the same
      geometrical scheme, the same subdivision of the geographical domain
      and accept the same inputs. 
      An additional set of parameters is used in 
      TRIGRS-P to specify the variability of the characteristics of the
      terrain.  Within each pixel, slopes are modeled as a~two-layer system
      consisting of a~lower saturated zone with a~capillary fringe above the
      water table, overlain by an unsaturated zone that extends to the
      ground surface. The water table and the (hypothetical) sliding surface
      are planar and parallel to the topographic surface. The geographical
      domain represented by an array of grid cells, coincides with the
      elements of a~digital elevation model (DEM) used to describe the
      topography of the study area (Figure~\ref{fig01}b).

\subsection{Deterministic approach: the TRIGRS code}\label{subsec:det_appr} 

      In the original approach coded in TRIGRS \citep{baum2008}, the
      stability of an individual grid cell is determined adopting the
      one-dimensional infinite-slope model \citet{taylor1948}. The model
      assumes that failure of a~grid cell occurs when the resisting forces
      $R$ acting on the sliding surface are less than the driving forces $S$
      \citep{wu1995,wyllie2004}. The ratio of the resisting $R$ and the
      driving $S$ forces gives by the factor of safety $F_{\mathrm{S}}$,
\begin{align}
F_{\mathrm{S}} = \frac{R}{S} = \frac{\tan \phi}{\tan \delta}
  +\frac{c - \psi\,\gamma_{\mathrm{w}} \tan \phi}{\gamma_{\mathrm{s}} z \sin \delta \cos \delta}\,,
  \label{eq01}
\end{align}
      where the internal friction angle $\phi$, the cohesion $c$, and the
      soil unit weight ${\gamma }_{\mathrm{s}}$ describe the material
      properties, $\gamma_{\mathrm{w}}$ is the groundwater unit weight,
      $\delta$ is the angle of the 
      planar slope, and $\psi$ is
      the pressure head (Figure~\ref{fig01}b, see Table~\ref{notation}). Failure occurs
      when $F_{\mathrm{S}}<1$. Solution of Equation~(\ref{eq01}) requires the
      \begin{figure*}[!htp]%
        \includegraphics[width=160mm]{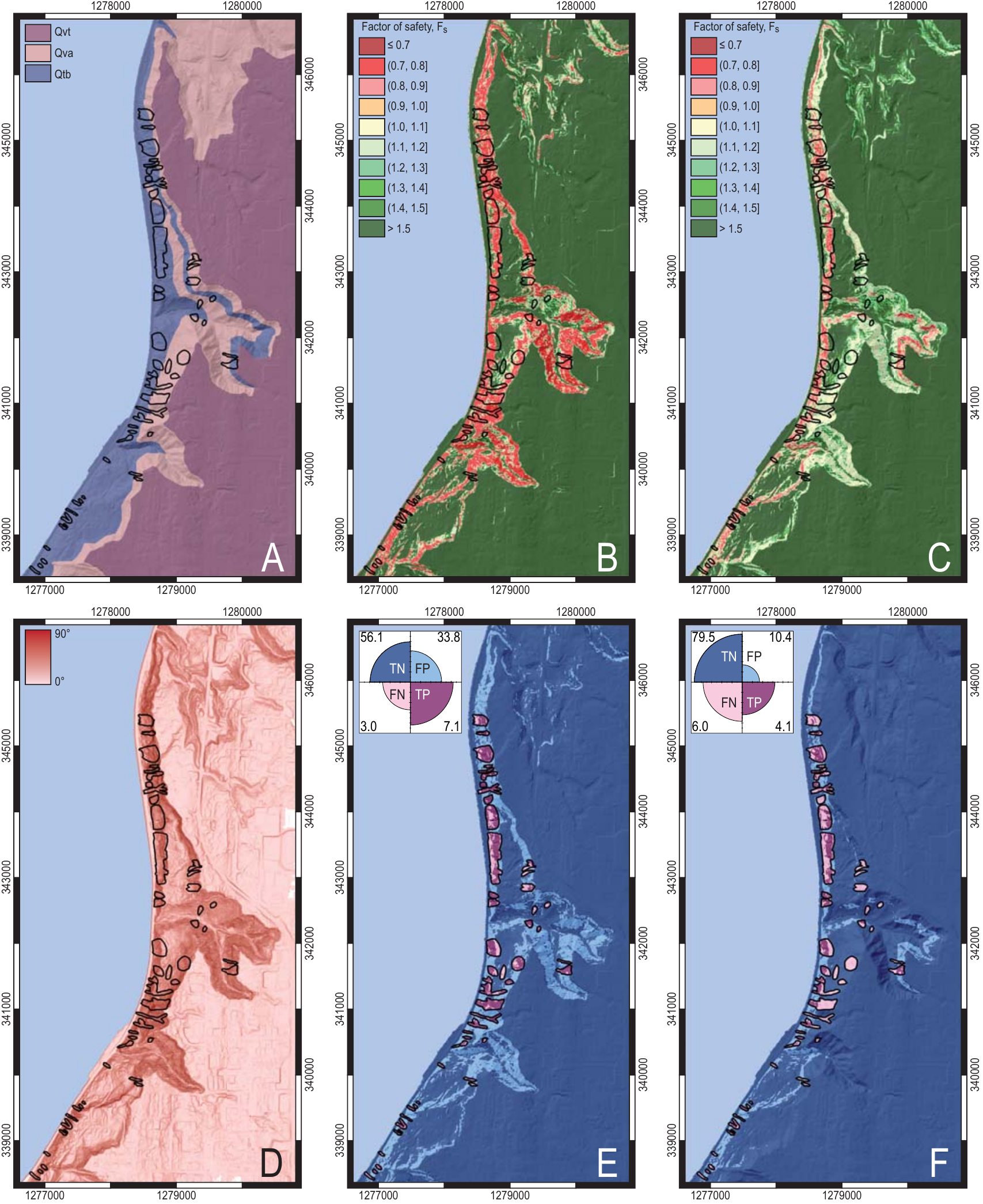}
        \caption{%
          Mukilteo study area; results obtained using the original
          TRIGRS code and input parameters of Table~\ref{tab01}.
          \textbf{(A)} Lithology map: Qtb, transition sediments, including the
          Lawton Clay ($1$ in Table~\ref{tab01});
          Qva, Advance outwash sand ($2$ in Table~\ref{tab01});
          Qvt, Glacial till ($3$ in Table~\ref{tab01}).
          \textbf{(B)} Factor of safety $F_{\mathrm{S}}$ obtained with saturated soil conditions;
          \textbf{(C)} $F_{\mathrm{S}}$ obtained with unsaturated soil conditions;
          \textbf{(D)} slope map;
          \textbf{(E)} map of correct assignments and model errors, within the saturated
          model; TP, True Positive; TN, True Negative; FP, False Positive;
          FN, False Negative;
          \textbf{(F)} as in \textbf{(E)}, for the unsaturated model.
          Black polygons show rainfall-induced landslides.}
        \label{fig03}%
      \end{figure*}%
      \begin{figure*}[!htp]%
        \includegraphics[width=160mm]{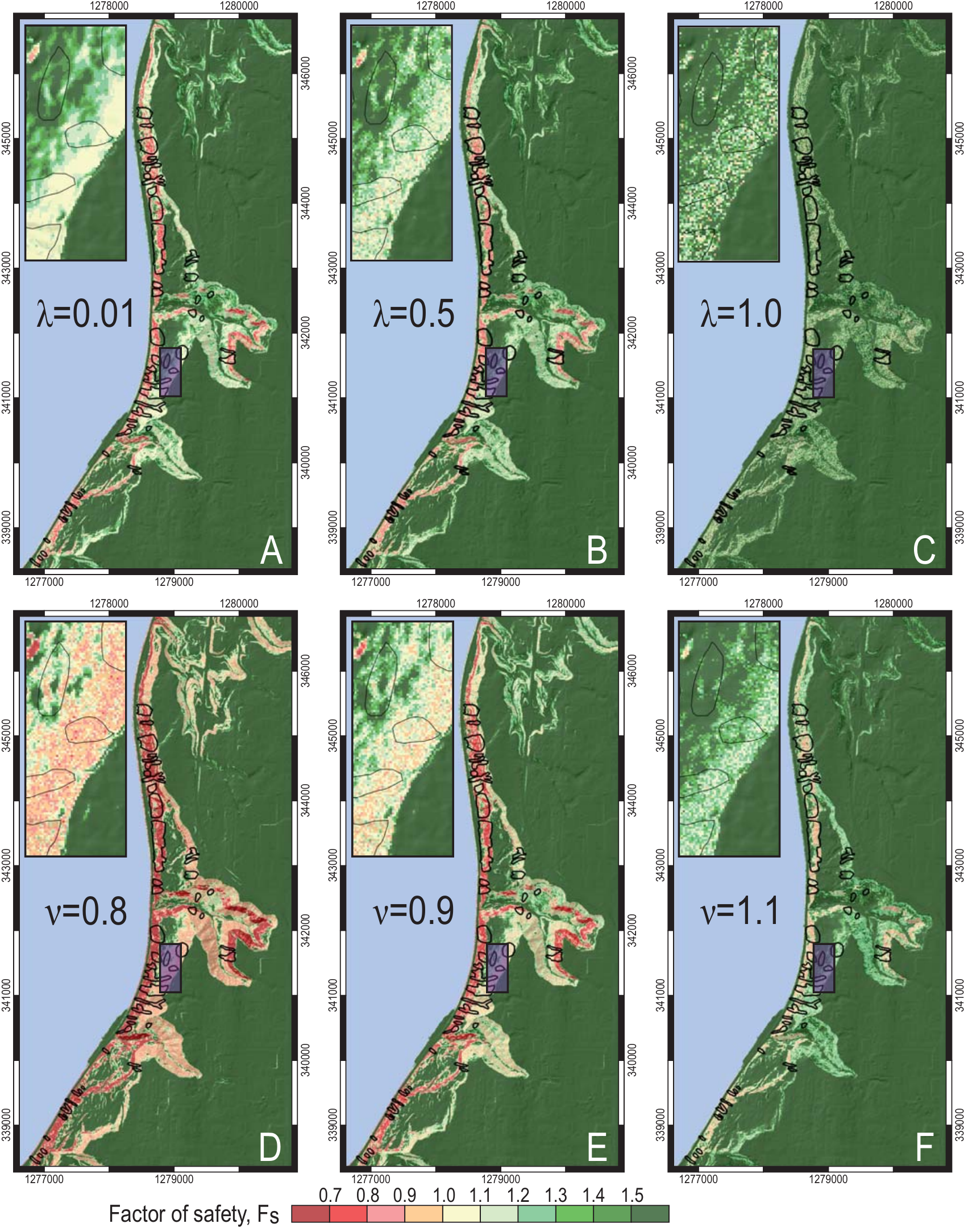}
        \caption{%
          Mukilteo study area. Maps showing factor of safety $F_{\mathrm{S}}$ obtained
          with the code presented in this work, TRIGRS-P, initialized with the
          same input parameters used for the same area and TRIGRS code, in Figure~\ref{fig03}, and
          with the following parameters for the random number generation:
          \textbf{(A)}~$\lambda =0.01$, $\nu =1.0$;
          \textbf{(B)}~$\lambda =0.5$, $\nu =1.0$;
          \textbf{(C)}~$\lambda =1.0$, $\nu =1.0$.
          \textbf{(D)}~$\lambda =0.5$, $\nu =0.8$;
          \textbf{(E)}~$\lambda =0.5$, $\nu =0.9$;
          \textbf{(F)}~$\lambda =0.5$, $\nu =1.1$.
          We performed 16 runs for each set of parameters. Black polygons show rainfall-induced
          landslides; the insets show the spatial variability of the factor of safety.
        }
        \label{fig04}%
      \end{figure*}%
      \begin{figure*}[!htp]%
        \includegraphics[width=160mm]{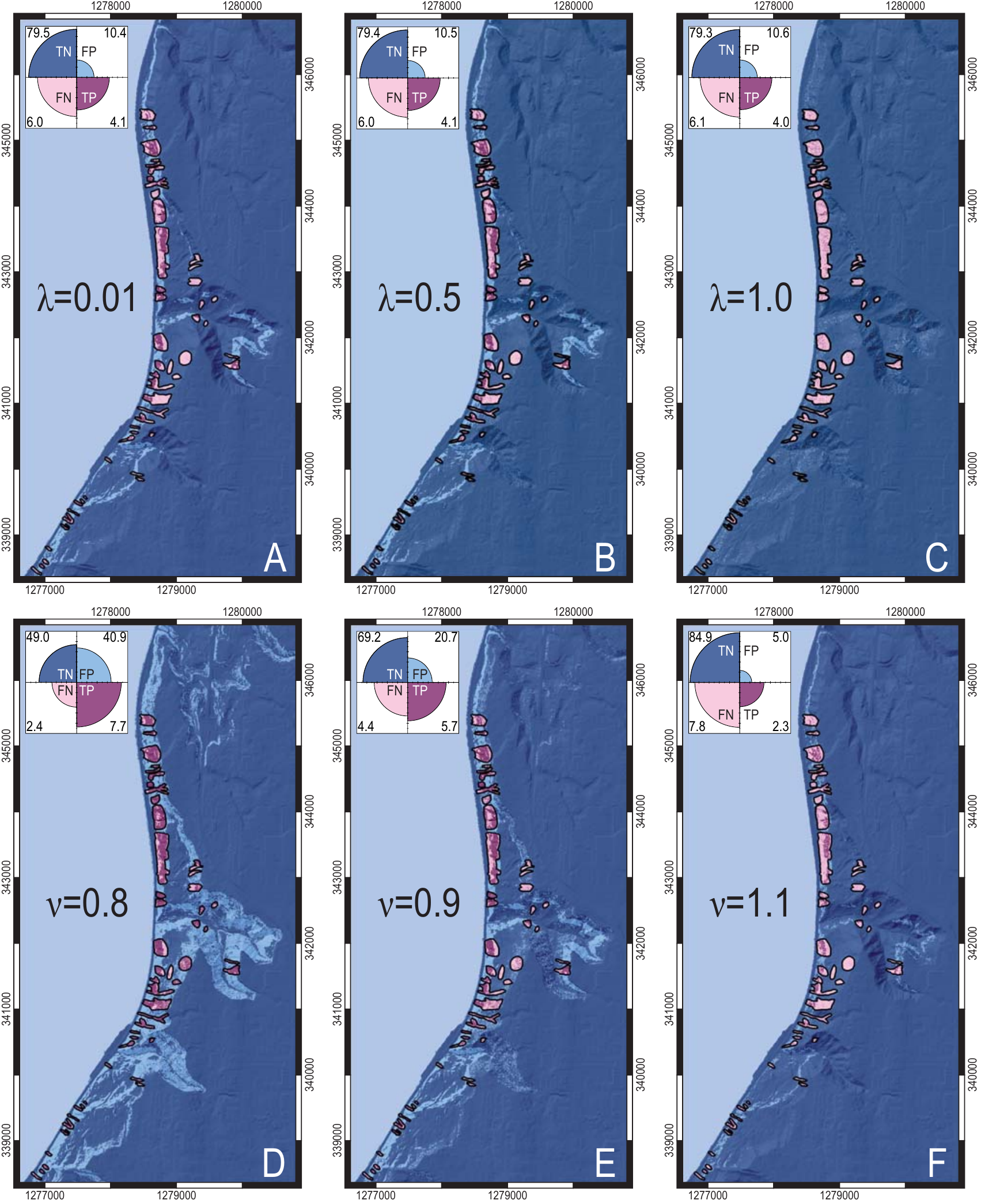}
        \caption{%
          Maps of correct assignments and model errors in the Mukilteo study area, obtained
          with the TRIGRS-P code with different sets of random input parameters.
          \textbf{(A)}~$\lambda =0.01$, $\nu =1.0$;
          \textbf{(B)}~$\lambda =0.5$, $\nu =1.0$;
          \textbf{(C)}~$\lambda =1.0$, $\nu =1.0$.
          \textbf{(D)}~$\lambda =0.5$, $\nu =0.8$;
          \textbf{(E)}~$\lambda =0.5$, $\nu =0.9$;
          \textbf{(F)}~$\lambda =0.5$, $\nu =1.1$.
          TP, True Positive; TN, True Negative; FP, False Positive; FN False
          Negative. In all maps, black polygons show rainfall-induced landslides in the study area.}
        \label{fig05}%
      \end{figure*}%
      computation of the pressure head $\psi$, which is governed by the
      \citet{richards1931} equation:
\begin{align}
\frac{\partial }{\partial z}\left[ K_z \left(\psi \right)
    \frac{\partial \left(\psi - z \right)}{\partial z}\right] =
  \frac{\partial \theta }{\partial t}\,,
  \label{eq02}
\end{align}
      where, $z$ is the slope-normal coordinate, $t$ is the time, $K_z$ is
      the vertical hydraulic conductivity that depends on the pressure head
      $\psi $, and $\theta$ is the volumetric water content
      (Figure~\ref{fig01}b). Equation~(\ref{eq02}) is solved in TRIGRS adopting the
      modeling scheme proposed by \citet{baum2008}.

      For saturated conditions, TRIGRS uses a~modified version of the
      analytical solutions of Equation~(\ref{eq02}) proposed by
      \citet{iverson2000}, for short term and for long-term rainfall
      periods. Again, the modification consists chiefly in the possibility
      of using a~complex rainfall history \citep{baum2008}. To linearize
      Equation~(\ref{eq02}), \citet{iverson2000} adopted a~normalization
      criterion using a~length scale ratio as follows:
\begin{align}
\varepsilon  = \sqrt {\frac{d^2_{\text{fp}} / D_0}{A / D_0}} = \frac{d_{\text{fp}}}{\sqrt{A}},
  \label{eq06}
\end{align}
      where $D_0$ is the maximum hydraulic diffusivity, $A$ is the
      contributing area that affects hydraulic pressure at the potential
      failure plane depth $d_{\text{fp}}$, and $d^{2}_{\text{fp}}/D_0$ and
      $A/D_0$ are the minimum time required for slope-normal
      ($d^{2}_{\text{fp}}/D_0$) and for slope-lateral ($A/D_0$) pore
      pressure transmission (see Table~\ref{notation}). Under the condition
      $\varepsilon \ll 1$, simplification of Equation~(\ref{eq02}) gives
      \citep{iverson2000}:
\begin{align}
\frac{\partial }{\partial z^*} \left[ K^*( \psi ) \left( \frac{\partial \psi^*}{\partial z^* } -
    z^* \right)\right] = 0\,,\,\,\,{\text{for}}\,\,\,t > \frac{A}{D_0}
  \label{eq07}
\end{align}
and
\begin{align}
\frac{\partial }{\partial z^*} \left[ K^* \left(\frac{\partial \psi *}{\partial z^*} -
    z^* \right)\right] = \frac{C (\psi)}{C_0} \frac{\partial \psi*}{\partial t^*}\,,\,\,\,{\text{for}}\,\,\,t\ll\frac{A}{D_0},
\label{eq08}
\end{align}
      where $\psi^* = \psi / d_{\text{fp}}$, $t^* = t D / A~$, and $z^* = z
      / \sqrt{d_{\text{fp}}}$.

      For unsaturated conditions, the code uses a~modified version of the
      analytical solution of Equation~(\ref{eq02}) proposed by
      \citet{srivastava1991}, for the case of one-dimensional, transient,
      vertical infiltration. The modification consists in the use of
      a~variable rainfall history (intensity, duration), allowing modeling
      of complex rainfall patterns \citep{baum2008}. Equation~(\ref{eq02})
      was linearized in \citet{srivastava1991}, who adopted the following
      exponential model \citep{gardner1958}:
\begin{align}
&K_z \left(\psi \right)=  K_{\mathrm{s}}\,e^{\alpha \widetilde{\psi}}\,;\\
  &\theta = \theta_{\mathrm{r}} + \left( \theta_{\mathrm{s}} - \theta_{\mathrm{r}}\right)\,e^{\alpha \widetilde{\psi }}\,,
  \label{eq03}
\end{align}
      where $K_{\mathrm{s}}$ is the saturated hydraulic conductivity,
      $\theta_{\mathrm{r}}$ is the residual water content,
      $\theta_{\mathrm{s}}$ is the saturated water content, and
      $\widetilde{\psi } = \psi - \psi_0$, $\psi_0 = -1/ \alpha$ is
      a~constant, with $\alpha $ the inverse of the vertical height of the
      capillary fringe above the water table \citep{savage2003,
      savage2004}. Substitution of Equation~(\ref{eq03}) into Equation~(\ref{eq02})
      leads to the partial differential equation:
\begin{align}
\frac{\alpha \left( \theta_{\mathrm{s}} - \theta_{\mathrm{r}} \right)}{K_{\mathrm{s}}} \frac{\partial K}{\partial t} =
  \frac{ \partial^2 K}{\partial z^2} - \alpha \frac{\partial K}{\partial z}\,.
  \label{eq04}
\end{align}
      Equation~(\ref{eq04}) is a~linear diffusion equation for which
      analytical solutions can be obtained using the Laplace, the Fourier,
      or the Green's function methods \citep{kevorkian1991}, once boundary
      conditions are specified, e.g.:
\begin{align}
&K(z,0)= I_{ZLT} - \left[ I_{ZLT}-K_{\mathrm{s}}e^{\alpha \psi_0} \right]e^{- \alpha z}\,;\\
  &K (0, t )= K_{\mathrm{s}}e^{\alpha {\psi }_0}
  \label{eq05}
\end{align}
      where $I_{ZLT}$ is the steady
      surface flux, which can
      be approximated by the average precipitation rate necessary to
      maintain the initial conditions in the days to months preceding an
      event \citep{baum2010}. When a~solution of Equation~(\ref{eq04}) is
      obtained, the pore pressure head $\psi$ can be calculated by inversion
      of Equation~(\ref{eq02}). Solutions of Equation~(\ref{eq04}) with the boundary
      conditions listed in Equation~(\ref{eq05})
are given in Appendix~A1.

      TRIGRS implements a~simple surface runoff routing scheme to disperse
      the excess water from the grid cells where rainfall intensity exceeds
      the local infiltration capacity \citep{hillel1982,baum2008}.

\subsection{Probabilistic approach: the TRIGRS-P code}\label{subsec:sto_appr} 

      In our extension of the TRIGRS code, we use the same model and
      equations as in the original code. The innovation consists of using
      probability distributions to model the slope material and hydrological
      properties, i.e. the values of the input parameters. The geometry of
      the slope ($\delta$) and the position of the sliding plane
      ($d_{\text{fp}}$) remain unchanged. The model parameters appearing in
      the equations described in Sect.~\ref{subsec:det_appr} are replaced by
      functions of random numbers, i.e.:
{\hack{\allowdisplaybreaks}
\begin{align}
&c=c ({\xi}_{\mathrm{c}} ), \; {\text{cohesion}};\nonumber\\
&  \phi =\phi ({\xi }_{\phi}), \; \mathrm{angle\,\,\,of\,\,\,internal\,\,\,friction};\nonumber\\
&      {\gamma }_{\mathrm{s}}=\gamma ({\xi }_{\gamma }), \; \mathrm{soil\,\,\,unit\,\,\,weight};\nonumber\\
&      D_0=D_0({\xi }_{D_0}), \; \mathrm{hydraulic\,\,\,diffusivity};\nonumber\\
&      K_{\mathrm{s}}=K_{\mathrm{s}}({\xi }_{K_{\mathrm{s}}}), \; \mathrm{saturated\,\,\,hydraulic\,\,\,conductivity};\nonumber\\
&      {\theta }_{\mathrm{r}}={\theta }_{\mathrm{r}}({\xi }_{\theta_{\mathrm{r}} }), \; \mathrm{residual\,\,\,water\,\,\,content};\nonumber\\
&      {\theta }_{\mathrm{s}}={\theta }_{\mathrm{s}}({\xi }_{\theta_{\mathrm{s}} }), \; \mathrm{saturated\,\,\,water\,\,\,content};\nonumber\\
&      \alpha=\alpha ({\xi }_{\alpha }), \; \mathrm{inverse\,\,\,height\,\,\,of\,\,\,capillary\,\,\,fringe}
\label{eq09}
\end{align}}
      where ${\xi}_i$ is a~random number, with the subscript \textit{i} used
      to specify a~different parameter, ${\xi}_{\mathrm{c}}$ for cohesion,
      ${\xi}_{\phi}$ for friction, etc., so that the parameters can be
      varied independently from each other.  Replacing the parameters listed
      in Equation~(\ref{eq09}) into Eqs.~(\ref{eq01}), (\ref{eq02}), (\ref{eq07}),
      (\ref{eq08}), and (\ref{eq04}), we obtain a~system of equations that
      are initialized with a~different, randomly chosen set of parameters at
      each run of TRIGRS-P.  The solution of the various scenarios for
      saturated or unsaturated conditions 
      are performed in the very same way as
      \begin{figure}[!htp]%
        \includegraphics[width=80mm]{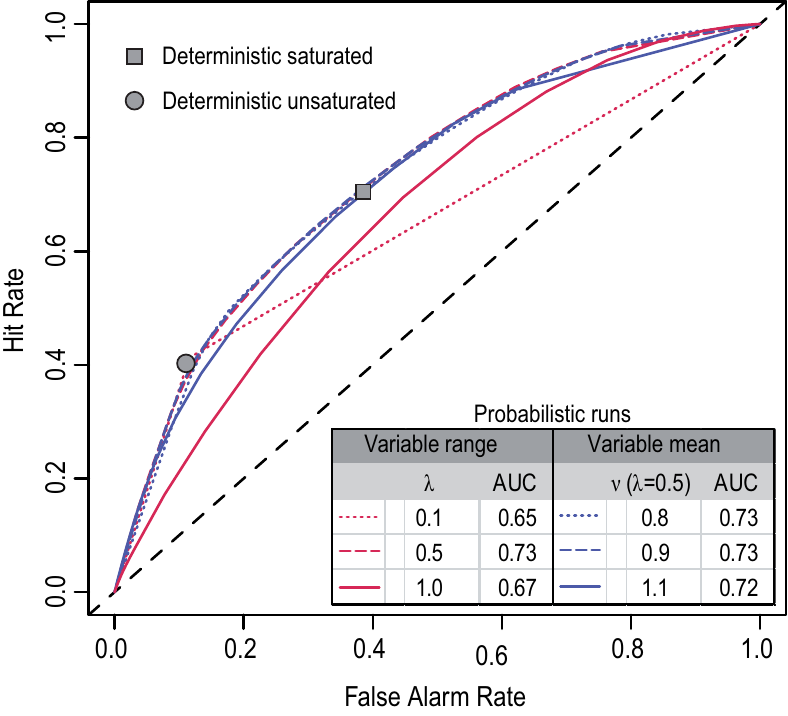}
        \caption{%
          The results of simulations for the Mukilteo study area, presented using
          ROC curves. The grey square and circle represent the results obtained using the
          original TRIGRS code with saturated and unsaturated initial conditions, respectively
          (Figure~\ref{fig03}b,~c); the curves correspond to the results obtained with the TRIGRS-P
          code, using the variability of input parameters shown in the inset as described in the
          text (Figures~\ref{fig04} and \ref{fig05}).}
        \label{fig06}%
      \end{figure}%
      in TRIGRS. The depth to the potential sliding plane $d_{\text{fp}}$
      was assumed to coincide with the soil depth, and was estimated by
      \citet{godt2008} and \citet{baum2010} using variations of the models
      proposed by \citet{derose1996} and by
      \citet{salciarini2006}. Additional choices for initial conditions and
      corresponding sources of uncertainties will be discussed in the
      following.

      We have implemented two probability density functions (pdf) for
      generating the modeling parameters: (i) the normal distribution
      function ${\mathcal N}$, and (ii) the uniform distribution function
      ${\mathcal U}$. If $\xi $ is a~standard normally distributed variable
      ${\mathcal N}{\text{(0,1}}$) with mean $\bar{\xi} =0 $ and standard
      deviation $\sigma = 1$, the variable $x = \overline{x}+ {\sigma }_x\xi
      $ is normally distributed with mean $\bar{x}$ and standard deviation
      ${\sigma}_x$, ${\mathcal N}(\bar{x},{\sigma }_x)$. Similarly, if $\xi
      $ is standard uniformly distributed ${\mathcal U} (0,1)$, the variable
      \begin{table}[!htb]
        \caption{Geotechnical parameters for the geological units cropping
          out in the Mukilteo area (Figure~\ref{fig03}a).
          $c$, cohesion; $D_{0}$, hydraulic
          diffusivity; $K_{\mathrm{s}}$, saturated hydraulic conductivity;
          $\theta_{\mathrm{s}}$, saturated water content; $\theta_{\text{r}}$, residual water
          content;  $\alpha $, inverse of capillary fringe.
          The friction angle $\phi$ has a~common value of 33.6{\degree} for
          the three geological units; units definitions are:
          1, Qtb; 2, Qva; 3, Qvt.}
                {
                  \begin{tabular}{lcccccc}
                    \tophline
                    Unit & $c$ & $D_0$ & $K_{\mathrm{s}}$ & $\theta_{\mathrm{s}}$ &
                    $\theta_{\mathrm{r}}$ & $\alpha$\\
                    & [kPa] &  [m$^2$\,s$^{-1}$] & [m\,s$^{-1}$] & -- & -- & [m$^{-1}$] \\
                    \middlehline
                    1 & 3.0 & $3.8\cdot 10^{-4}$ & $1.0\cdot  10^{-4}$ & $0.40$ & $0.06$ & $10$ \\
                    2 & 3.0 & $5.0\cdot 10^{-6}$ & $1.0\cdot  10^{-7}$ & $0.40$ & $0.10$ & ~\,$2$ \\
                    3 & 8.0 & $8.3\cdot 10^{-6}$ & $1.0\cdot 10^{-6}$ & $0.45$ & $0.10$ & ~\,$5$ \\
                    \bottomhline
                \end{tabular}}
                \belowtable{%
                }\label{tab01}
      \end{table}
      \begin{figure}[!htp]%
        \includegraphics[width=80mm]{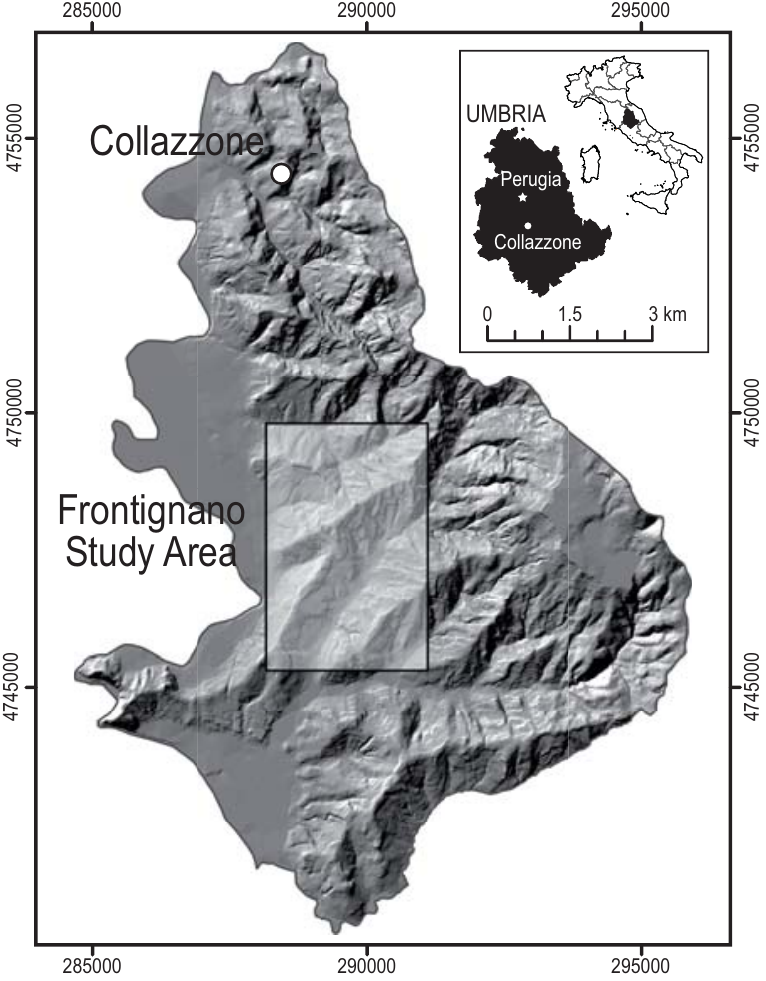}
        \caption{%
          Map showing location of the Frontignano study area, Umbria, central Italy.
          The area is located inside the larger Collazzone area \citep{guzzetti2006a, guzzetti2006b}.}
        \label{fig07}%
      \end{figure}%
      $y = y_a +(y_b - y_a )\xi$ is uniformly distributed in the range
      $[y_a,y_b]$, ${\mathcal U}(y_a,y_b)$. The advantage of using these
      expansions is that their deterministic limits are obtained for
      ${\sigma }_x\to 0$ and for $\lambda \equiv y_b -y_a\to 0$,
      respectively.

      In this work, we calculated the stability conditions in the modeling
      domain for a~given set of variables describing the slope materials
      properties ($\phi$, $c$, ${\gamma}_{\mathrm{s}}$, $K_{\mathrm{s}}$,
      $D_0$, ${\theta }_{\mathrm{r}}$, ${\theta }_{\mathrm{s}}$) obtained by
      sampling randomly from the uniform distribution only.  There is
      a~conceptual difference between the two distributions for distributed
      landslide probabilistic modeling. Adoption of the Gaussian distribution
      requires that the investigator has determined (e.g., through
      sufficient field tests or laboratory experiments) the uncertainty and
      \mbox{measuring} errors associated with the parameters. The mean and the
      standard deviation of the Gaussian 
      \begin{table}[!hbp]
        \caption{Estimators of model performance for saturated and unsaturated soil
          calculated with the original TRIGRS code, for the Mukilteo study area.
          TPR, True Positive Rate; FPR, False Positive Rate; ACC, Accuracy;
          PPV, Precision.}
                {
                  \begin{tabular}{lcccc}
                    \tophline
                    Model type & TPR & FPR & ACC & PPV\\
                    \middlehline
                    Saturated & $0.71$ & $0.38$ & $0.63$ & $0.17$ \\
                    Unsaturated & $0.41$ & $0.12$ & $0.84$ & $0.28$ \\
                    \bottomhline
                \end{tabular}}
                \belowtable{%
                }\label{tab02}
      \end{table}
      distribution define unambiguously
      the uncertainty. Use of the uniform distribution implies that the
      investigator only knows the possible (or probable) range of variation
      of the parameters, ignoring the internal structure of the
      uncertainty. We consider the Gaussian distribution more appropriate to
      predict rainfall-induced landslides in small areas where sufficient
      field and laboratory tests were performed to characterize the physical
      properties of the geological materials, and the uniform distribution
      best suited in the investigation of large areas where information on
      the geo-hydrological properties is limited. Further, we consider use
      of the Gaussian distribution best suited to investigate how errors in
      the parameters propagate and affect the modeling results, provided
      that the errors are known. Conversely, use of the uniform distribution
      allows for investigating how the uncertainty in the model parameters
      affects the model results. The sensitivity of the extended model to
      the random variation of model parameters has been explored by running
      16 independent simulations, each with a~different set of input
      parameters while keeping unchanged, and equal to the run performed
      with the original fixed-input TRIGRS model, the terrain morphology
      ($\delta$) and rainfall history.

\section{Deterministic vs. probabilistic approach}\label{sec:model_res}

      We tested the performance of the new probabilistic version of the
      numerical code, TRIGRS-P $2.0$, against the original TRIGRS code,
      version~2.0 \citep{baum2008}, in two study areas. The first test was
      conducted in the Mukilteo study area, near Seattle, WA, USA
      (Figure~\ref{fig02}). This is the same geographical area where
      \citet{godt2008} and \citet{baum2010} demonstrated the use of TRIGRS
      in a~broad geographical setting. The second test was performed in the
      Frontignano study area, Perugia, Italy (Figure~\ref{fig07}). This is
      part of the Collazzone geographical area where
      \citet{guzzetti2006a,guzzetti2006b} have investigated the hazard posed
      by shallow landslides using multivariate classification methods.
      \subsection{Mukilteo study area}\label{subsec:mukilteo_area}

      The three square kilometre study area is located along the eastern
      side of the Puget Sound, about $15\,\unit{km}$ north of Seattle, WA,
      USA (Figure~\ref{fig02}). In this area, rainfall is the primary trigger
      of landslides. Slope failures are typically shallow (less than three
      meters thick), and involve the sandy colluvium and the weathered
      glacial deposits mantling the coastal bluffs
      \citep{galster1991,baum2000}. The climate of the Seattle area is
      characterized by a~pronounced seasonal precipitation regime with
      a~winter maximum, and $3/4$ of the annual precipitation falling from
      November to April \citep{church1974}. Storms that trigger shallow
      landslides in Seattle are generally of long duration (more than 24\,h)
      and of moderate intensity \citep{godt2006}. Three geological units
      crop out in the area \citep{minard1982} (Figure~\ref{fig03}a) including,
      from older to younger: (i) transition sediments, comprising the Lawton
      Clay (Qtb), (ii) advance outwash sand (Qva), and (iii) glacial till
      (Qvt). The mechanical and hydrological properties of the materials in
      the three geological units are known through field tests and
      laboratory experiments \citep{lu2006,godt2006,godt2008}, and are
      summarized in Table~\ref{tab01}.

\subsubsection{Predictions with the deterministic approach}
\label{subsubsec:mukilteo_det}

      For modeling purposes, the topography of the area was described by
      a~$6\,\unit{ft} \times 6\,\unit{ft}$ ($1.83\,\unit{m} \times 1.83\,\unit{m}$) DEM
      obtained through airborne laser-swath mapping
      \citep{haugerud2003}. Initial conditions for infiltration were
      prescribed as zero pressure head at the depth of the lower boundary of
      colluvium. This is in agreement with field observations
      \citep{baum2005a,schulz2007,godt2008}.  A~constant rainfall intensity
      $I = 4.5\,\unit{mm\,h^{-1}}$ for a~period of 28\,\unit{h} was used to force
      slope instability, for a~cumulative event rainfall $E = 126\,\unit{mm}$. The
      adopted rainfall history represents a~limit case of the rainfall
      intensity-duration conditions that have resulted in landslides in the
      Mukilteo area in the winter 1996--1997 \citep{godt2008,baum2010}.
      Figure~\ref{fig03} shows the results of the runs with deterministic
      input, for saturated (Equation~\ref{eq07}, Figure~\ref{fig03}b) and for
      unsaturated (Equation~\ref{eq08}, Figure~\ref{fig03}c) conditions. For the
      mechanical and hydrological properties of the geological materials
      ($\phi$, $c$, ${\gamma }_{\mathrm{s}}$, $K_{\mathrm{s}}$, $D_0$,
      ${\theta }_{\mathrm{r}}$, ${\theta }_{\mathrm{s}}$) we considered the
      values listed in Table~\ref{tab01}.

      In order to test the model prediction skills i.e., the ability of the
      model to forecast the known distribution of rainfall-induced
      landslides \citep{guzzetti2006a}, the two geographical distributions
      of the factor of safety $F_{\mathrm{S}}$ were compared to a~landslide
      inventory showing slope failures triggered by rainfall in the winter
      1996--1997 \citep{baum2000,godt2008}, displayed by black lines in
      Figure~\ref{fig03}.  For the comparison, all grid cells with
      $F_{\mathrm{S}} <1$ were considered unstable (i.e., potential
      landslide) cells. Four-fold plots and maps showing the geographical
      distribution of the correct assignments and the model errors
      (Figure~\ref{fig03}e,~f) are used to summarize and display the
      comparison. Four-fold plots are graphical representations of
      contingency tables (or confusion matrices), and show the fraction (or
      number) of true positives (TP), true negatives (TN), false
      positives (FP), and false negatives (FN)
      \citep{fawcett2006,rossi2010}. In our analysis TP is the percentage
      of cells with observed landslides, which are predicted as unstable by
      the model; similarly, TN is the percentage of cells without
      landslides predicted as stable by the model. Correspondingly, FP
      (FN) are the percentage of predicted unstable (stable) cells without
      (with) observed landslides. We will refer to both TP and TN as
      correct assignment in the following, while FP and FN are model
      errors.  To further quantify the performance of the deterministic
      forecasts, different metrics were computed (Table~\ref{tab02}),
      \begin{figure*}[!htp]%
        \includegraphics[width=160mm]{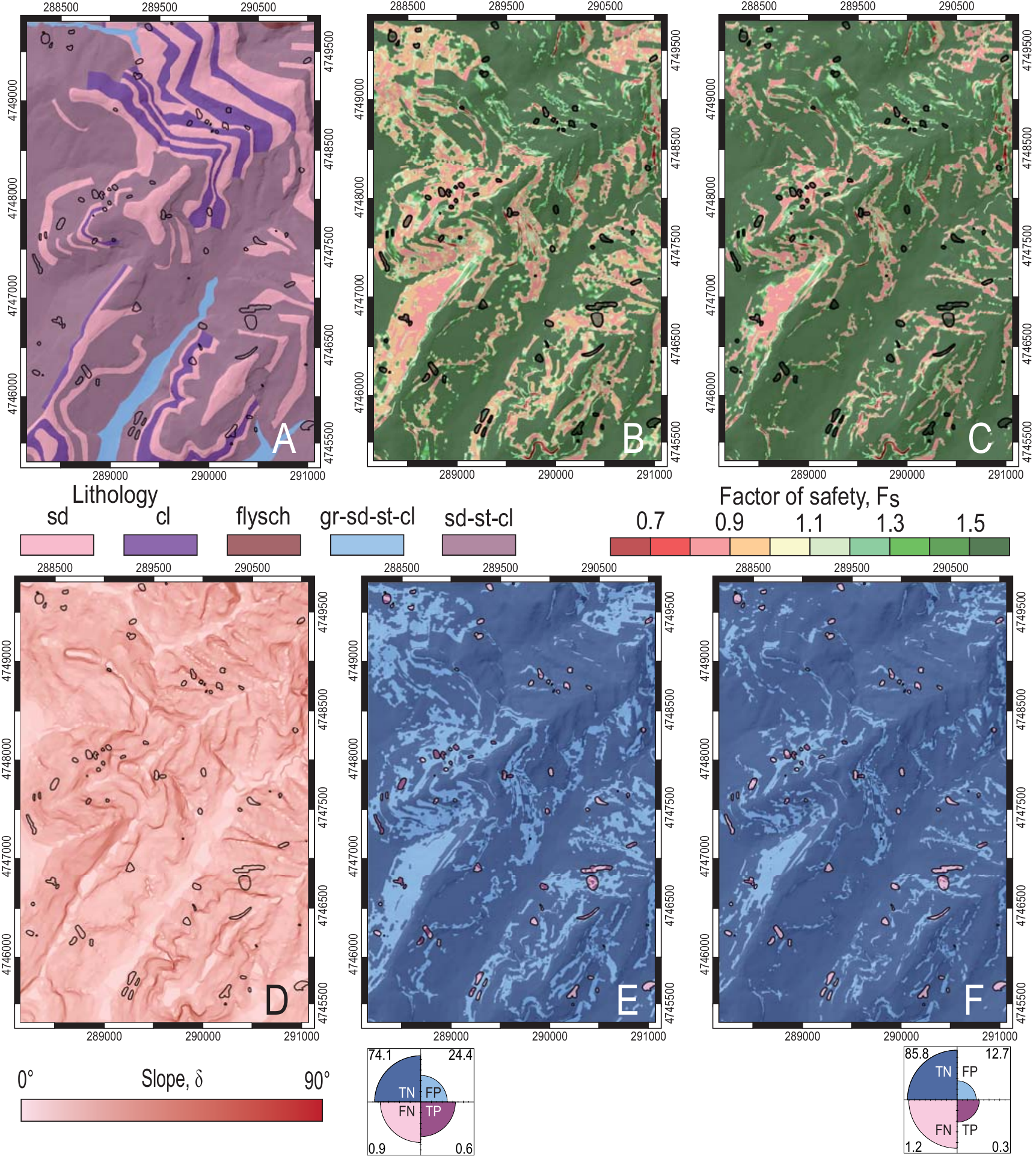}
        \caption{%
          As in Figure~\ref{fig03}, but for the Frontignano study area.
          The results have been obtained using the original
          TRIGRS code with input parameters listed in Table~\ref{tab05}.
          \textbf{(A)}~Lithological map: sand ($1$ in Table~\ref{tab05}); clay ($2$ in Table~\ref{tab05});
          flysch deposits ($3$ in Table~\ref{tab05}); gravel, sand, silt, and
          clay ($4$ in Table~\ref{tab05}); sand, silt, and clay ($5$ in Table~\ref{tab05}).
          \textbf{(B)}~Factor of safety $F_{\mathrm{S}}$ obtained with saturated soil conditions;
          \textbf{(C)}~$F_{\mathrm{S}}$ obtained with unsaturated soil conditions;
          \textbf{(D)}~slope map;
          \textbf{(E)}~map of correct assignments and model errors, within the saturated model;
          TP, True Positive; TN, True Negative; FP, False Positive; FN, False
          Negative;
          \textbf{(F)}~as in \textbf{(E)}, for the unsaturated model.
          Black polygons show rainfall-induced landslides.}
        \label{fig08}%
      \end{figure*}%
      \begin{figure*}[!htp]%
        \includegraphics[width=160mm]{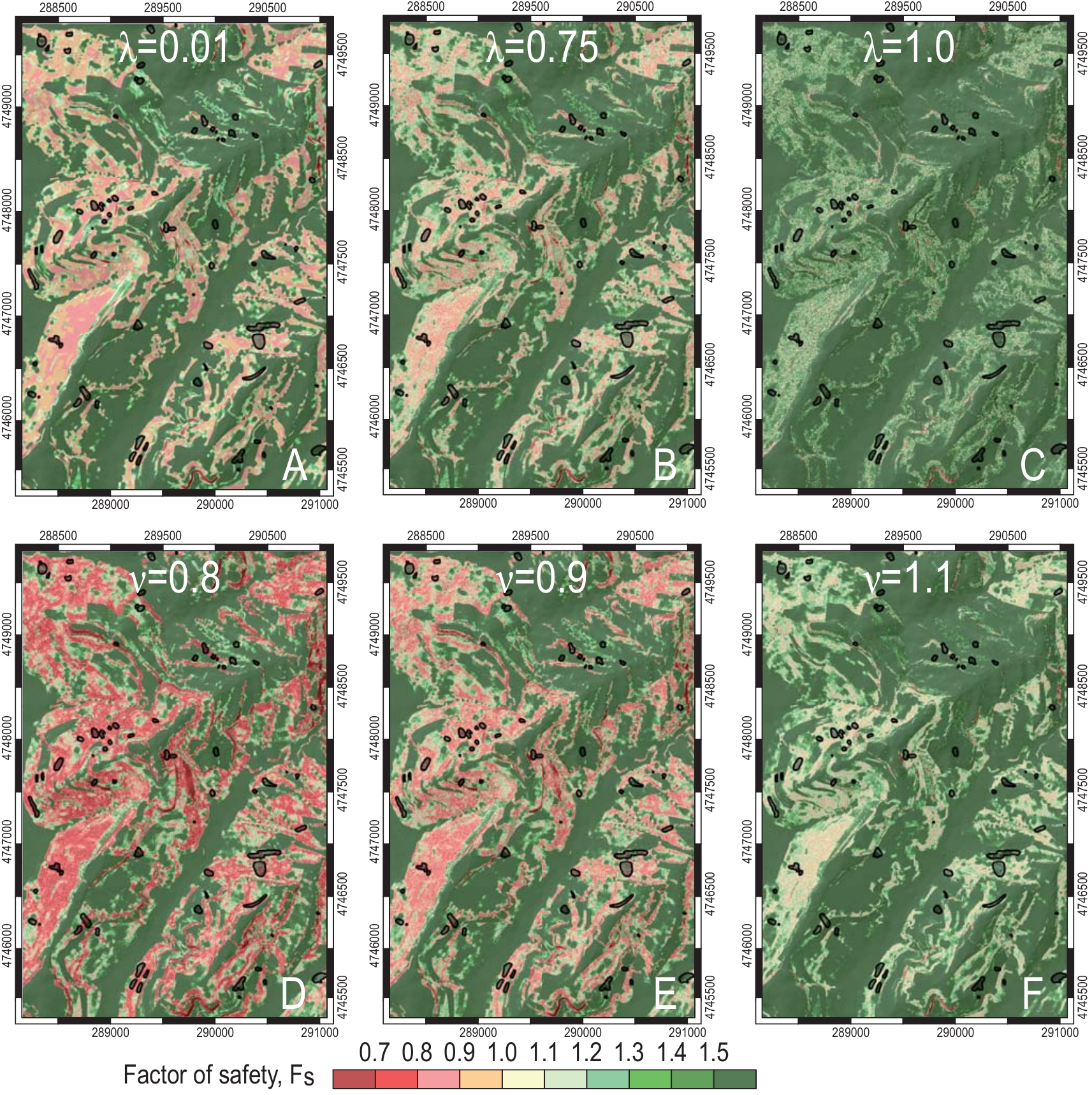}
        \caption{%
          Frontignano study area. Maps of the factor of safety $F_{\mathrm{S}}$ obtained
          within the probabilistic approach of TRIGRS-P, with the following values of range of
          variation of input parameters:
          \textbf{(A)}~$\lambda =0.01$, $\nu =1.0$; \textbf{(B)}~$\lambda =0.75$, $\nu =1.0$; \textbf{(C)}~$\lambda =1.0$, $\nu =1.0$.
          \textbf{(D)}~$\lambda =0.75$, $\nu =0.8$; \textbf{(E)}~$\lambda =0.75$, $\nu =0.9$; \textbf{(D)}~$\lambda =0.75$, $\nu =1.1$.
          We performed 16 runs for each set of parameters. In all maps, black polygons show rainfall-induced
          landslides in the study area.}
        \label{fig09}%
      \end{figure*}%
      \begin{figure*}[!htp]%
        \includegraphics[width=160mm]{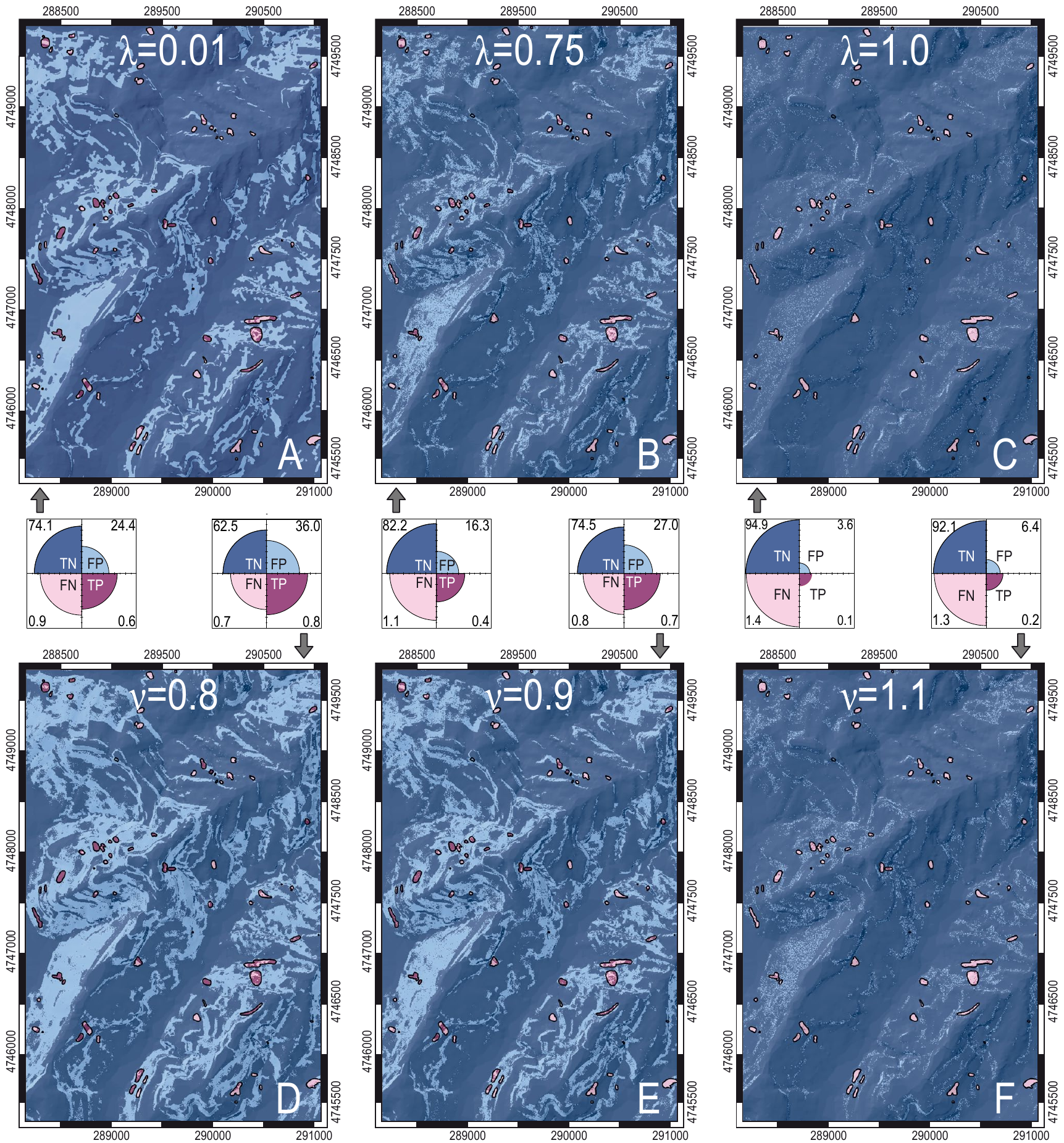}
        \caption{%
          Frontignano study area. Maps of the factor of correct assignments
          and model error obtained within the probabilistic approach of TRIGRS-P, with the following
          values of range of variation of input parameters:
          \textbf{(A)}~$\lambda =0.01$, $\nu =1.0$; \textbf{(B)}~$\lambda =0.75$, $\nu =1.0$; \textbf{(C)}~$\lambda =1.0$,
          $\nu =1.0$. \textbf{(D)}~$\lambda =0.75$, $\nu =0.8$; \textbf{(E)}~$\lambda =0.75$, $\nu =0.9$; \textbf{(D)}~$\lambda =0.75$, $\nu =1.1$. TP, True Positive; TN, True Negative; FP, False Positive;
          FN False Negative. In all maps, black polygons show rainfall-induced landslides in the study area.}
        \label{fig10}%
      \end{figure*}%
      \begin{figure}[!htp]%
        \includegraphics[width=80mm]{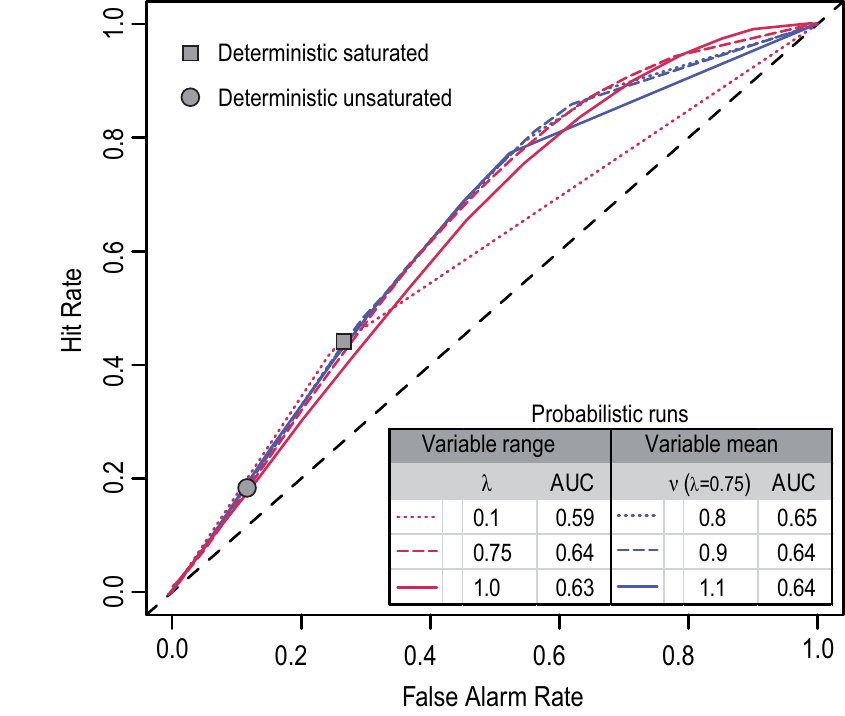}
        \caption{%
          Frontignano study area; ROC plots corresponding to
          the runs with fixed input parameters (Figure~\ref{fig08}b,~c)
          and with the probabilistic approach with random input parameters
          (Figures~\ref{fig09} and \ref{fig10}).}
        \label{fig11}%
      \end{figure}%
      including the True Positive Rate (sensitivity, or hit rate) ${\text{TPR}}
      ={\text{TP}}/({\text{TP}}+{\text{FN}})$, the True Negative Rate (specificity) 
      ${\text{TNR}} = {\text{TN}}/({\text{FP}}+{\text{TN}})$,
      the False Positive Rate ($1$ -- specificity, or false alarm rate) ${\text{FPR}}
      = {\text{FP}}/({\text{FP}}+{\text{TN}})$, the Accuracy
      ${\text{ACC}} = ({\text{TP}}+{\text{TN}})/({\text{TP}}+{\text{FN}}+{\text{FP}}+{\text{TN}})$, and the
      Precision ${\text{PPV}} = {\text{TP}}/({\text{TP}}+ {\text{FP}})$ \citep{fawcett2006,baum2010}.

\subsubsection{Predictions with the probabilistic approach}
\label{subsubsec:mukilteo_sto}

      Based on the comparison of the results discussed in the previous
      section, for the probabilistic modeling we used only the unsaturated soil
      conditions, and we exploited the same geomorphological information
      (i.e., the same DEM) and the same rainfall forcing input (i.e., $4.5\,\unit{mm\,h^{-1}}$ of rain 
      for a~28-h period) used for the previous
      runs. For the mechanical and hydrological properties of the geological
      materials ($\phi$, $c$, ${\gamma }_{\mathrm{s}}$, $K_{\mathrm{s}}$,
      $D_0$, ${\theta }_{\mathrm{r}}$, ${\theta }_{\mathrm{s}}$) we
      considered the values listed in Table~\ref{tab01}, used in the
      previous paragraph as fixed inputs of the model, as mean values of
      uniformly distributed variables ${\mathcal U}(y_a,y_b)$,
      \begin{table}[!hbp]
        \caption{Estimators of model performance of the results obtained with the TRIGRS-P
          code in the Mukilteo study area. In this case we change the ranges of variation
          of the model parameters $\lambda$, with fixed mean values of the model parameters
          $\nu= 1.0$. TPR, True Positive Rate; FPR, False Positive Rate; ACC, Accuracy;
          PPV, Precision; AUC, Area Under the ROC curve.}
                {
                  \begin{tabular}{lccccc}
                    \tophline
                    $\lambda $ & TPR & FPR & ACC & PPV & AUC\\
                    \middlehline
                    $0.01$ & $0.41$ & $0.11$ & $0.84$ & $0.29$ & $0.65$ \\
                    $0.10$ & $0.41$ & $0.11$ & $0.84$ & $0.29$ & $0.70$ \\
                    $0.50$ & $0.40$ & $0.12$ & $0.83$ & $0.28$ & $0.73$ \\
                    $0.75$ & $0.34$ & $0.11$ & $0.83$ & $0.26$ & $0.71$ \\
                    $1.00$ & $0.09$ & $0.04$ & $0.87$ & $0.22$ & $0.67$ \\
                    \bottomhline
                \end{tabular}}
                \belowtable{%
                } \label{tab03}
      \end{table}
      where $y_a$ and $y_b$ are the lower and upper limits of the uniform
      distribution determining the range of variation of each parameter. In
      our simulations, the range of variation of the individual parameters
      has been chosen as a~fraction of the mean value of each
      variable. A~range of variation $\lambda=0.01$, 0.10 and 1.00
      correspond to a~variation of $1\,\unit{\%}$, $10\,\unit{\%}$ and
      $100\,\unit{\%}$ around the mean value of the variable,
      respectively. Note that the case with $\lambda =0.01$ allows the
      various input parameters to vary in a~very limited range, and it can
      be seen as a~test of our code: the original TRIGRS results with fixed
      input parameters should be obtained.

      We performed two sets of runs. In the first set, the mean values of
      the mechanical and hydrological parameters (Table~\ref{tab01}) were
      kept constant, and the range of variation of the individual parameters
      was modulated using $\lambda =0.01$, 0.1, 0.5, 1.0. In the second
      set, a~fixed range of variation for the individual parameters was
      selected, $\lambda =1.0$, and the mean value of the parameters was
      modified (shifted) by $\nu =0.2$, $0.4, \dots, 1.0$, 2.0.  Note that
      when $\nu =1.0$, no shift of the mean value is performed. In each
      test, the same range of variation $\lambda $ and the same shift of the
      mean value $\nu $ were applied to all the parameters. The
      simplification was adopted to reduce the time required to perform
      multiple runs. The results are shown in Figure~\ref{fig04}: (i) for the
      first set of runs, i.e. for fixed mean values of the model parameters
      and changing ranges of variation of the individual parameters,
      $\lambda =0.01$ (Figure~\ref{fig04}a), $\lambda =0.5$
      (Figure~\ref{fig04}b), and $\lambda =1.0$ (Figure~\ref{fig04}c), and (ii)
      for the second set of runs, i.e. for a~fixed range of variation
      $\lambda =1.0$, and shifting the mean value of the model parameters by
      $\nu =0.8$ (Figure~\ref{fig04}d), $\nu =0.9$ (Figure~\ref{fig04}e), and
      $\nu =1.1$ (Figure~\ref{fig04}e). For the second set of runs, results
      obtained for $\nu < 0.8$ and for $\nu > 1.1$ are not shown in
      Figure~\ref{fig04}. For $\nu < 0.8$ the number of unconditionally
      unstable cells was unrealistically large, and for $\nu > 1.1$ the
      model performance decreased rapidly (see next paragraph).  We used 16
      runs for each set, resulting in 16 different maps of the factor of
      safety, which were used to evaluate the performance of the
      probabilistic approach.  The results are shown in Figure~\ref{fig05}.
      For the same runs of Figure~\ref{fig04}, the maps show the geographical
      distribution of the correct assignments (TP,~TN), the model
      errors (FP,~FN), and the corresponding four-fold
      plots. Tables~\ref{tab03} and \ref{tab04} list metrics that quantify
      \begin{table}[!hbp]
        \caption{As in Table~\ref{tab03}, but with fixed ranges of variation of
          and with varying the mean values the model parameters, for the Mukilteo
          area.}
                {
                  \begin{tabular}{lccccc}
                    \tophline
                    $\nu $ & TPR & FPR & ACC & PPV & AUC\\
                    \middlehline
                    $0.8$ & $0.77$ & $0.46$ & $0.56$ & $0.16$ & $0.73$ \\
                    $0.9$ & $0.57$ & $0.23$ & $0.75$ & $0.22$ & $0.73$ \\
                    $1.0$ & $0.40$ & $0.12$ & $0.83$ & $0.27$ & $0.73$ \\
                    $1.1$ & $0.23$ & $0.06$ & $0.87$ & $0.31$ & $0.72$ \\
                    \bottomhline
                \end{tabular}}
                \belowtable{%
                }\label{tab04}
      \end{table}
      \begin{figure*}[!htp]%
        \includegraphics[width=160mm]{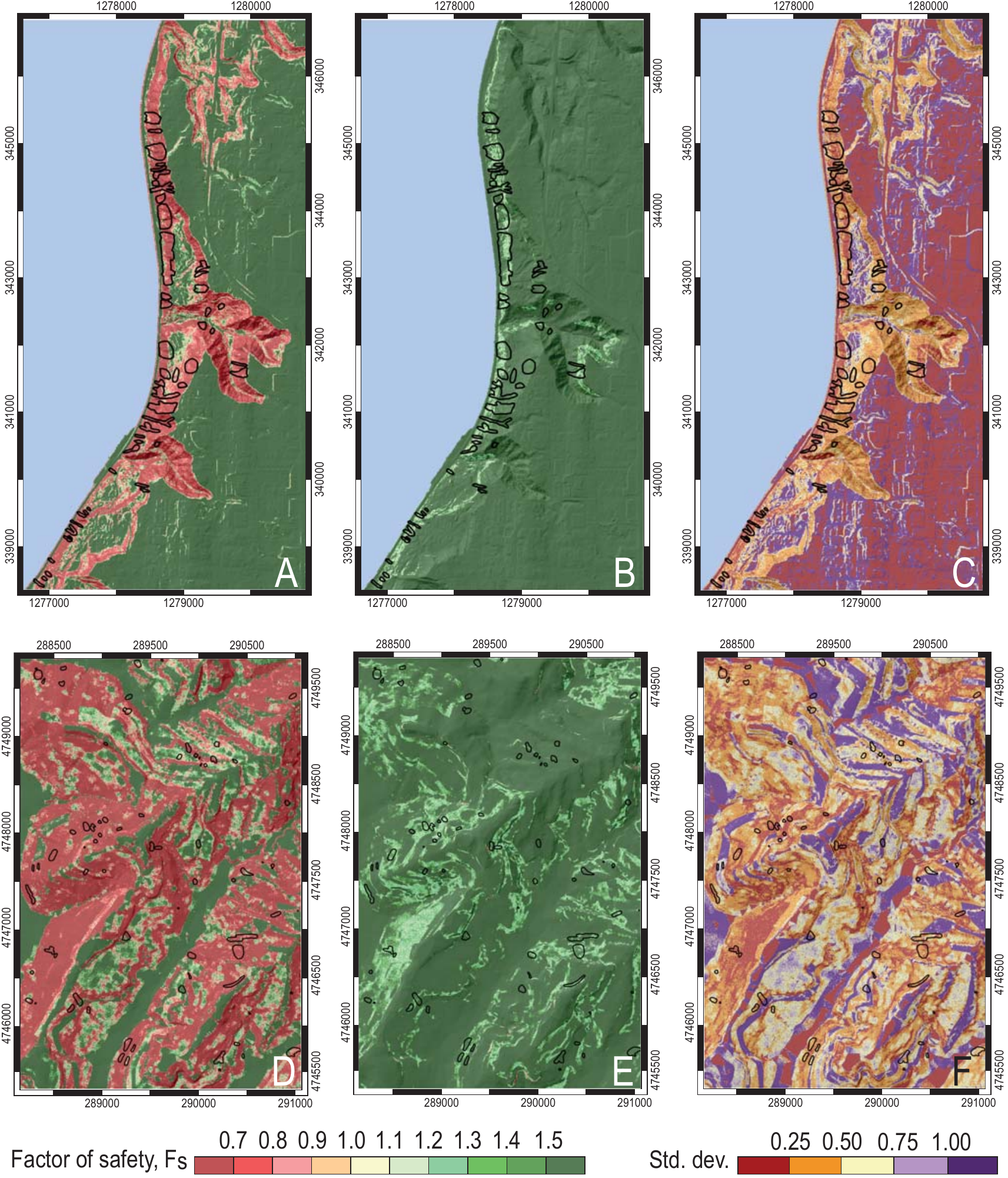}
        \caption{%
          Maps showing the minimum (left column), maximum (centre column),
          and standard deviation (right column), of the factor of safety $F_{\mathrm{S}}$ for the set of
          16 simulation runs using the TRIGRS-P code. Maps are shown for the Mukilteo (upper row)
          and the Frontignano (lower row) study areas.}
        \label{fig12}%
      \end{figure*}%
      the performance of the probabilistic approach.

\subsubsection{Analysis and discussion}
\label{subsubsec:mukilteo_compare}

      Inspection of the results of the deterministic (Figure~\ref{fig03}) and
      the probabilistic (Figures~\ref{fig04} and \ref{fig05}) models, and of
      their forecasting skills (Figure~\ref{fig06}, Tables~\ref{tab02}--\ref{tab04}), allows for general
      considerations. Figure~\ref{fig06} shows a~Receiver Operating
      Characteristics (ROC) plot \citep{fawcett2006}, defined by the false
      alarm rate FPR and the hit rate TPR, plotted on the x- and y-axes, respectively. 
      In the ROC space, a~point located in the upper
      left corner represents a~perfect prediction (${\text{TPR}} = 1$ and ${\text{FPR}} = 0$),
      and points along the diagonal line for which ${\text{TPR}} = {\text{FPR}}$ represent
      random predictions. An acceptable prediction requires ${\text{TPR}}/{\text{FPR}} > 1$
      \citep{fawcett2006}. In Figure~\ref{fig06}, two separate points show the
      predictive performance of the two runs with deterministic inputs, for
      saturated (Figure~\ref{fig03}b) and for unsaturated (Figure~\ref{fig03}c)
      conditions each of them producing a~single pair FPR-TPR and
      a~unique geographical distribution of the factor of safety
      $F_{\mathrm{S}}$. Analysis of Figures~\ref{fig03} and \ref{fig06}, and
      of Table~\ref{tab02} indicates that the model prepared considering the
      soil unsaturated conditions (Figure~\ref{fig03}c) performed better than
      the model prepared considering saturated conditions
      (Figure~\ref{fig03}b). The larger value of the TPR$/$FPR ratio is
      a~measure of the better predicting performance of the unsaturated
      model (${\text{TPR}}/{\text{FPR}} = 3.46$, Figure~\ref{fig06}), compared to the saturated
      model (${\text{TPR}}/{\text{FPR}} = 1.87$, Figure~\ref{fig03}), despite a~lower TPR
      value (${\text{TPR}} = 0.42$ vs. ${\text{TPR}} = 0.71$, Table~\ref{tab02}). This is in
      agreement with previous work of \citet{godt2008} and~\citet{baum2010}.

      Within the two deterministic models, the one using the unsaturated
      soil conditions (Figure~\ref{fig03}c,~f) performed better than the model
      that used the saturated soil conditions (Figure~\ref{fig03}b,~e). The
      saturated model predicted a~significantly larger fraction of the study
      area as unstable, mainly where terrain gradient exceeded
      $15{\degree}$. This resulted in \mbox{a~considerably} larger number of true
      positives (TP, 7.1\,\unit{\%} vs. 4.1\,\unit{\%}), but also
      a~significantly larger number of false positives (FP, 33.8\,\unit{\%} vs. 10.4\,\unit{\%}) and a~correspondingly
      significantly lower number of true negatives (TN, 56.1\,\unit{\%}
      vs. 79.5\,\unit{\%}). In other words, the saturated deterministic
      model (Figure~\ref{fig03}b) was more pessimistic than the unsaturated
      deterministic model (Figure~\ref{fig03}c). This is well represented in
      Figure~\ref{fig06}, where a~reduction of the false positive rate from
      0.38 to 0.12 results in a~reduction of the hit rate from 0.71 to
      0.41 (Table~\ref{tab02}). The subsequent runs with probabilistic
      input were obtained assuming unsaturated soil water conditions. The
      results of the unsaturated probabilistic models (Figure~\ref{fig04}) were
      similar to the results of the corresponding unsaturated deterministic
      model (Figure~\ref{fig03}c). This is a~significant result, confirming
      that treating the uncertainty associated with the model parameters
      with a~probabilistic approach has not significantly changed the model
      results, which have remained consistent. Availability of multiple
      model outputs for each run allowed preparing ROC curves to measure
      quantitatively the predictive performance of the probabilistic models
      \citep{fawcett2006}.  Since multiple values of $F_{\mathrm{S}}$ are
      available for each pixel in the modeling domain, we can calculate the
      frequency of stability condition of each pixel. We attribute to this
      frequency the meaning of a~probability and compare it with a~given
      threshold.  Modulation of the classification threshold allows us to
      obtain different FPR and TPR values, which can be used to
      construct a~ROC curve \citep{fawcett2006}. In Figure~\ref{fig06} two
      sets of ROC curves are shown using different colors. The red curves
      show the performances of the first set of runs, for $\lambda =0.01$,
      $\lambda =0.5$, and $\lambda =1.0$, with $\nu =1.0$, and the blue
      curves show the performances of the second set of runs, for $\nu
      =0.8$, $\nu =0.9$, and $\nu =1.1$, with $\lambda =0.5$. To construct
      the ROC curves, several probability thresholds were used, from $0.1$
      to $0.9$ by $0.1$ steps. The area under the ROC curve AUC is taken
      as a~quantitative measure of the performance of the classification. If
      ${\text{AUC}} = 0.5$, a~classification is poor and indistinguishable from
      a~random classification, whereas a~perfect classification has ${\text{AUC}} =
      1$ \citep{fawcett2006,rossi2010}.

      Inspection of Figures~\ref{fig05} and~\ref{fig06}, 
      and of Table~\ref{tab03}, suggests that an increase in the range of variation
      of the model parameters (from $\lambda =0.01$ to $\lambda =1.0$),
      corresponding to \mbox{a~significantly} larger degree of uncertainty in the
      parameters, resulted in similar individual performance indices, but
      significantly larger values of the area under the ROC curve, AUC. In
      our experiment, the increase in the range of variation changed the
      performance index from ${\text{AUC}} = 0.65$ (for $\lambda =0.01$) to ${\text{AUC}} =
      0.73$ (for $\lambda =0.5$), with an increase of performance of
      16\,\unit{\%}.  A~further increase of the range of variation to
      $\lambda =1.0$, a~possibly unrealistic range of variation for some of
      the modeling parameters, has resulted in a~value of ${\text{AUC}} = 0.67$,
      decreasing the model performance. Modulation of the mean value of the
      parameters, using $\nu =0.8$, $\nu =0.9$, and $\nu =1.1$, resulted in
      better results (larger AUC values) for $\lambda =0.5$ than for
      $\lambda =0.01$.  Moreover, the TPR, FPR, PPV and ACC metrics
      did not change significantly when the range of variation $\lambda $ of
      the model parameters were modified, and remained similar to the values
      obtained with the deterministic models, for $\lambda \le 0.5$. We
      conclude that, in the Mukilteo study area, these metrics are not
      sensitive to introduction of the probabilistic determination of the
      model parameters. Second, the AUC showed a~positive correlation with
      the range of variation in the model parameters.

      In the probabilistic runs, a~positive correlation was observed between
      the range of variation $\lambda $ and the fraction of unconditionally
      unstable cells i.e., the grid cells that have $F_{\mathrm{S}} <1$ even
      in dry conditions when no rainfall is increasing pore pressure and
      slope instability.  For the first set of runs, the fraction of
      unconditionally unstable cells was $0\,\unit{\%}$ for $\lambda =0.01$,
      $0.3\,\unit{\%}$ for $\lambda =0.5$, and $0.7\,\unit{\%}$ for $\lambda
      =1.0$. Moreover, a~negative correlation was observed between $\nu$,
      the width of shift in the mean value of the modeling parameters, and
      the fraction of unconditionally unstable cells. For the second set of
      runs, the fraction of unconditionally unstable cells was less than
      $5.0\,\unit{\%}$ for $\nu \ge 0.8$, and was $0\,\unit{\%}$ for $\nu >
      1.0$, independent of the range of variation of the parameters.

\subsection{Frontignano study area}\label{subsec:frontig_area}

      The Frontignano area is located in central Umbria, Italy, about
      $25\,\unit{km}$ south of Perugia, in the Collazzone area
      (Figure~\ref{fig07}). In this area, landslides are caused primarily by
      rainfall and rapid snowmelt
      \citep{cardinali2000,guzzetti2006a,guzzetti2006b,fiorucci2011}. Multiple
      deep-seated and shallow slides were identified in the area through the
      visual interpretation of multiple sets of aerial photographs and
      very-high resolution satellite images, and field surveys.

      The shallow failures are typically less than three meters thick, and
      involve the soil and the colluvium mantling the slopes. Soils range in 
      thickness from a~few decimetres to more than one meter; they have
      a~fine to medium texture, and exhibit a~xeric moisture regime, typical
      of the Mediterranean climate. In central Umbria, precipitation is most
      abundant in October and November, with a~mean annual rainfall in the
      period 1921--2001 exceeding 850\,mm. In the study area, terrain
      is hilly, and the lithology and the attitude of bedding planes control
      the morphology of the slopes. Gravel, sand, clay, travertine, layered
      sandstone and marl, and thinly layered limestone, crop out in the area
      \citep{cardinali2000,guzzetti2006a,guzzetti2006b}.

\subsubsection{Predictions with the deterministic approach}\label{subsubsec:frontig_det}

      For modeling purposes, the topography of the Frontignano study area
      was described by a~$5\,\unit{m} \times  5\,\unit{m}$ DEM obtained interpolating
      $5$-m contour lines shown on $1:10\,000$ scale topographic base
      maps \citep{guzzetti2006a,guzzetti2006b}. Slope in the area ranges
      from $0{\degree}$ to $62{\degree}$, with an average value of
      $10{\degree}$ and a~standard deviation of $5.6{\degree}$
      (Figure~\ref{fig08}d). The mechanical and hydrological properties of the
      five soil types cropping out in the area (Figure~\ref{fig08}a) were
      determined through laboratory tests and searching the literature
      \citep[see, e.g.][and references
      therein]{shafiee2008,feda1995,lade2012} on the geotechnical properties
      ($\phi$, $c$, ${\gamma }_{\mathrm{s}}$, $K_{\mathrm{s}}$, $D_0$,
      ${\theta }_{\mathrm{r}}$, ${\theta }_{\mathrm{s}}$) of the same or
      similar sediments in Umbria, Italy (listed in Table~\ref{tab05}). As
      for the Mukilteo area, the depth to the hypothetical sliding plane
      $d_{\text{fp}}$ was assumed to coincide with the soil depth, which was
      estimated using the model proposed by \citet{derose1996}. To calibrate
      the soil depth model, we exploited field observations indicating that
      the depth of the shallow landslides in the study area is
      $d_{\text{fp}}<3$\,m, and that shallow landslides are most
      abundant where terrain gradient is in the range $7{\degree} \le \delta
      \le 20{\degree} $. Initial depth to the water table was set to
      a~fraction of the depth to the failure plane, ${d_{\mathrm{w}}= 0.85
      d}_{\text{fp}}$. 
      Since the depth of the water table is an important initial condition for the model, 
      we decided to use a long rainfall period, starting from an almost dry initial condition 
      and reaching a realistic depth of the water table during the storm. 
      We further decided not to set the water table to the maximum soil depth 
      to consider the fact that the simulation is intended to be representative 
      of typical winter conditions, when landslides occur in both study areas, 
      and when the soil always contains some amount of water. 
      We tested different rainfall histories, and adopted
      \mbox{a~forcing} rainfall that produced shallow landslides in the area in the
      periods January--May 2004, October--December 2004, and
      October--December 2005
      \citep{guzzetti2009,fiorucci2011}. Specifically, we used a~rainfall
      history composed of a~$4$-week initial rainfall period characterized
      by a~constant mean rainfall intensity $I = 0.36\,\unit{mm\,h^{-1}}$, for
      a~cumulative rainfall $E = 242\,\unit{mm}$, followed by a~$60$-min
      rainfall period characterized by a~high rainfall intensity $I= 90\,\unit{mm\,h^{-1}}$, for a~cumulative rainfall $E = 90\,\unit{mm}$. Results for the
      saturated (Equation~\ref{eq07}) and the unsaturated (Equation~\ref{eq08})
      modeling conditions are shown in Figures~\ref{fig08}b,~c, respectively.

     To test the model performance, the geographical distribution 
      of the factor of safety $F_{\mathrm{S}}$ predicted by TRIGRS 
      were compared to the known distribution of rainfall-induced landslides
      mapped in the same area in the periods January to May 2004, October
      to December 2004, and October to December 2005. The landslides
      were mapped through reconnaissance fieldwork and the visual
      interpretation of high-resolution satellite images
      \citep{guzzetti2009,fiorucci2011}, and are shown with black lines in
      Figure~\ref{fig08}.  For the comparison, all grid cells with
      $F_{\mathrm{S}} <1$ were considered unstable (i.e., landslide)
      cells. As for the Mukilteo test case, four-fold plots
      (Figure~\ref{fig08}e,~f) and derived metrics (Table~\ref{tab06}), ROC
      plots (Figure~\ref{fig11}), and maps showing the geographical
      distribution of the correct assignments and the model errors
      (Figure~\ref{fig08}e,~f) were used to summarize and measure the
      comparison.

      Inspection of Figures~\ref{fig08} and~\ref{fig11}, and analysis of
      Table~\ref{tab06}, suggests that the saturated and the unsaturated
      models produce very similar results. This is different from the result
      obtained in the Mukilteo area, where the unsaturated model performed
      better than the saturated model. In the Frontignano area, the
      unsaturated model (Figure~\ref{fig08}c) resulted in a~better forecasting
      accuracy (ACC, 0.86 vs. 0.75), but in a~reduced TPR to FPR
      ratio ($1.4$ vs. $1.7$). We maintain that the model prepared
      considering the saturated conditions (Figure~\ref{fig08}b) performed
      slightly better than the model obtained considering the unsaturated
      conditions (Figure~\ref{fig08}c).

\subsubsection{Predictions with the probabilistic approach}\label{subsubsec:frontig_sto}

      The mechanical and hydrological properties of the geological materials
      ($\phi$, $c$, ${\gamma }_{\mathrm{s}}$, $K_{\mathrm{s}}$, $D_0$,
      ${\theta }_{\mathrm{r}}$, ${\theta }_{\mathrm{s}}$) in the Frontignano study area 
      were chosen as listed in Table~\ref{tab05} (also used as input of the
      original TRIGRS model, in the previous paragraph) as mean values of
      uniformly distributed variables ${\mathcal U}(y_a,y_b)$. To
      be consistent with the approach adopted in Mukilteo, we performed two
      sets of parametric analyses, varying the range ($\lambda $) and the
      mean value ($\nu $) of the model parameters. The maps in
      Figure~\ref{fig09} show the factor of safety $F_{\text{S}}$ calculated
      \begin{figure}[!htp]%
        \includegraphics[width=80mm]{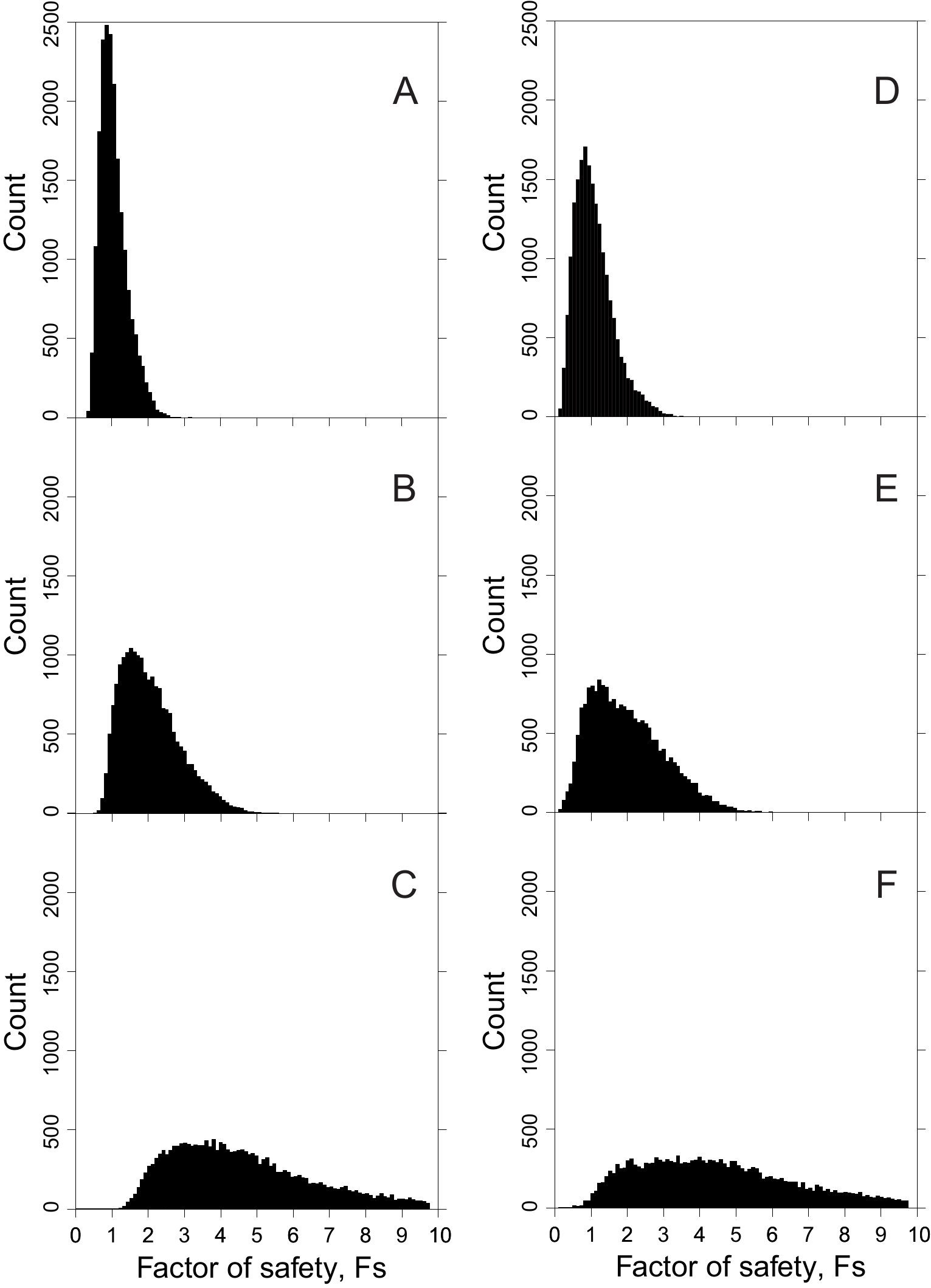}
        \caption{%
          Histograms showing the distribution of the values of the $F_{\mathrm{S}}$ for the Mukilteo
          (left, \textbf{A}, \textbf{B}, \textbf{C}) and the Frontignano (right, \textbf{D}, \textbf{E}, \textbf{F}) study areas. \textbf{(A)}~and \textbf{(D)} for subsets of $1000$
          grid cells with $0<\overline{F}_{\mathrm{S}} \le 1.5$. \textbf{(B)} and \textbf{(E)} for subsets of $1000$ grid cells with
          $1.5<\overline{F}_{\mathrm{S}}\le 3$. \textbf{(C)} and \textbf{(F)} for subsets of 1000 grid cells with such that $\overline{F}_{\mathrm{S}}>3$.}
        \label{fig13}%
      \end{figure}%
      for: (i) fixed mean values of the model parameters $\nu =1.0$, and
      changing ranges of variation of the individual parameters, $\lambda
      =0.01$ (Figure~\ref{fig09}a), $\lambda =0.75$ (Figure~\ref{fig09}b), and
      $\lambda =1.0$ (Figure~\ref{fig09}c), and (ii) a~fixed range of
      variation $\lambda =0.75$, and shifting the mean value of the model
      parameters by $\nu =0.8$ (Figure~\ref{fig09}d), $\nu =0.9$
      (Figure~\ref{fig09}e), and $\nu =1.1$ (Figure~\ref{fig09}f). As in the
      previous case, $\lambda =0.01$ corresponds to a~very small range of
      variability of the parameters, and provides the same results. 
      For $\nu =1.0$, no shift in the mean values of the model parameters is 
      performed. The degree of accuracy of the two sets of runs for the
      Frontignano area is shown in Figure~\ref{fig10}, for the same
      models shown in Figure~\ref{fig09}. The maps show the geographical
      distribution of the correct assignments (TP,  TN), the model
      errors (FP, FN), and the corresponding four-fold
      plots. Tables~\ref{tab07} and \ref{tab08} list metrics that quantify
      the performance of the runs. The performance of the probabilistic models
      is further analysed in Figure~\ref{fig11} by two sets of ROC curves,
      shown using different colours; red curves for the case of variable
      range $\lambda $, and blue curves for the case of a~variable mean
      $\nu$. In the same plot, the grey circle shows the predicting
      performance of the saturated model (Figure~\ref{fig08}b), and the grey
      square the performance of the unsaturated model (Figure~\ref{fig08}c)
      both run with fixed input parameters.

\subsubsection{Analysis and discussion}\label{subsubsec:frontig_compare}

      Inspection of the results of the fixed input runs (Figure~\ref{fig08}),
      the runs with input parameter sampled from a~suitable probability
      distribution, (Figures~\ref{fig09} and \ref{fig10}), and of their
      ability to forecast the spatial distribution of known landslides
      (Figure~\ref{fig11}, Tables~\ref{tab06}--\ref{tab08}),
      allows for 
      considerations that are similar to those discussed
      for the Mukilteo study area (see Sect.~3.1.3), with a~few
      differences. In the Frontignano area, the saturated and the
      unsaturated models provided 
      nearly equivalent results, with the
      saturated model considered marginally superior primarily because of
      the reduced value of the TPR to FPR ratio. From a~statistical
      point of view, given the reduced fraction of landslide area in
      Frontignano ($1.5\,\unit{\%}$) compared to Mukilteo
      ($4.2\,\unit{\%}$), the spatial prediction of landslides in
      Frontignano was more difficult than in Mukilteo. From a~physical point
      of view, modeling the stability conditions in low gradient terrain is
      very sensitive to the initial conditions, which are uncertain and
      difficult to determine spatially. The runs with variable input
      parameters confirm the slightly poorer geographical predictive
      performance of the adopted physical framework in Frontignano, compared
      to Mukilteo (Tables~\ref{tab03} and \ref{tab04} vs. Tables~\ref{tab07}
      and \ref{tab08}). Taking the area under the ROC curve (AUC) as the
      metric to compare the models, one can readily see that runs for the
      Mukilteo area resulted in $0.65 \le {\text{AUC}} \le 0.73$, and for the
      Frontignano area exhibited $0.59 \le {\text{AUC}} \le 0.65$. In other words,
      the ``worst'' result for Mukilteo (${\text{AUC}} = 0.65$, for $\nu =1.0$ and
      $\lambda =0.01$) has the same overall spatial predictive performance
      of the ``best'' result for Frontignano (${\text{AUC}} = 0.65$, for $\nu =0.8\
      or\ 0.9$ and $\lambda =0.75$). In the Frontignano area, despite
      a~lower ``absolute'' performance (i.e., when compared to Mukilteo),
      adoption of a~probabilistic approach improved the spatial forecasting
      skills. Again, taking AUC as a~metric to compare the models, values of
      this metric increased from ${\text{AUC}} = 0.59$ (for $\nu =1.0$ and $\lambda
      =0.01$), to ${\text{AUC}} = 0.65$ (for $\nu =0.8$ or 0.9 and
      $\lambda=0.75$). This is a~non-negligible improvement of about
      $10\,\unit{\%}$. The result confirms that adoption of a~probabilistic
      framework to the distributed modeling of shallow landslides results in
      improved spatial forecasts. 
      
      The result further corroborates the 
      finding that modeling the natural uncertainty (and poor understanding)
      of the mechanical and hydrological variables results in better spatial
      landslide predictions of the locations of rainfall-induced landslides
      (see insets in Figure~\ref{fig04}). First, the TPR, FPR, PPV, and AUC 
      metrics did not change significantly when the range of variation
      $\lambda$ of the model parameters was changed. These metrics remained
      similar to the values obtained with the fixed input model, confirming
      \begin{figure*}[!htp]%
        \includegraphics[width=140mm]{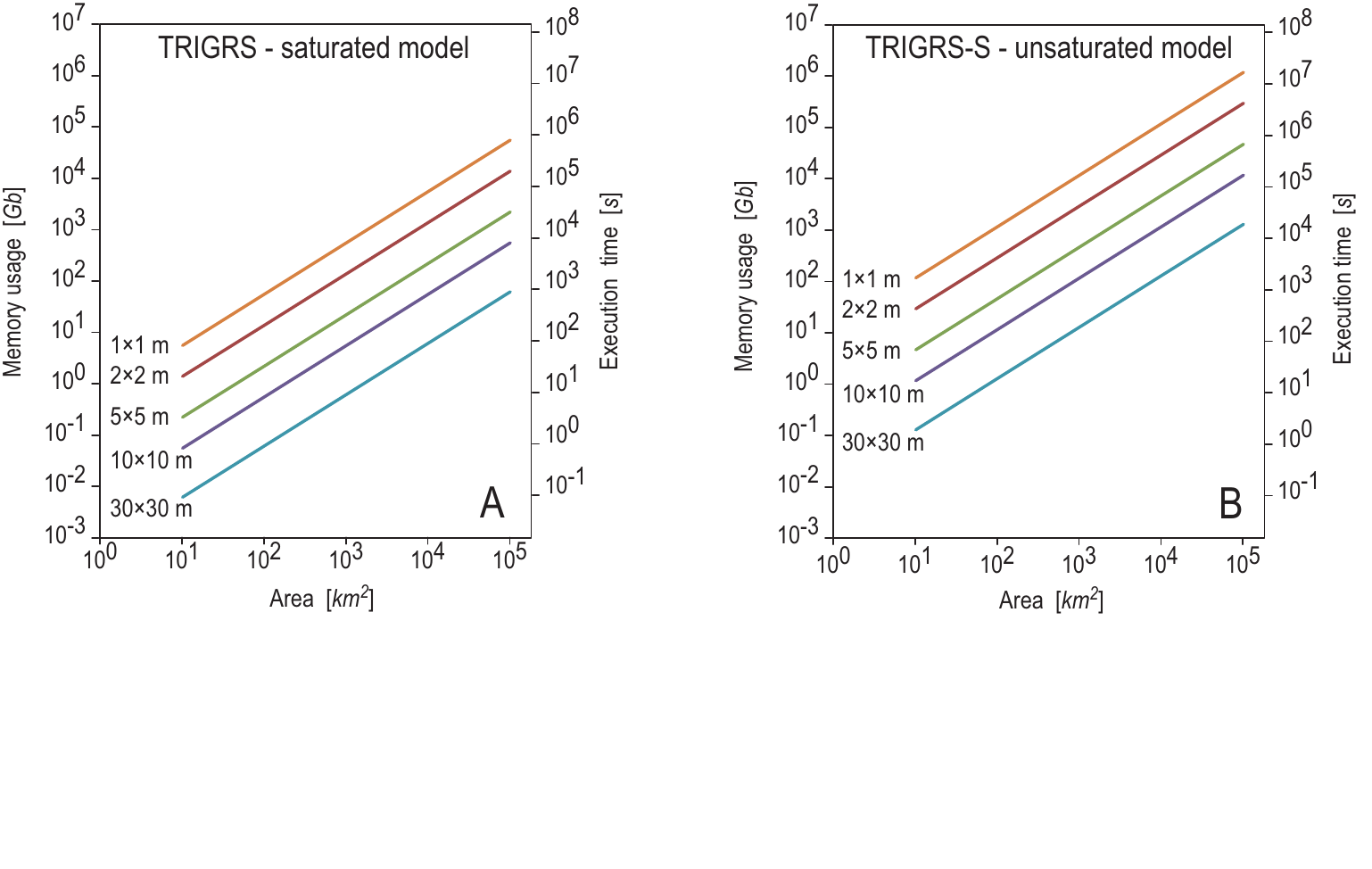}
        \vskip -2cm
        \caption{%
          Estimated memory usage (left y-axis) and execution times
          (right y-axis) for \textbf{(A)} the TRIGRS code (saturated model), and \textbf{(B)} a~set of 16
          runs of the TRIGRS-P code, for areas of different extent, and for grid cells of
          different spaptial resolutions.}        
        \label{fig14}%
      \end{figure*}%
      that they are not sensitive to differences between probabilistic
      framework runs with random variations of parameters and runs with
      fixed parameters. Second, the area under the ROC curve AUC confirmed
      its positive correlation with the range of variation in the model
      parameters $\lambda$, in support of the probabilistic approach. Third,
      the positive correlation between the range of variation $\lambda$ and
      the fraction of unconditionally unstable cells, and the negative
      correlation between the shift in the mean value of the modeling
      parameters $\nu$ and the fraction of unconditionally unstable cells,
      were both confirmed.  \section{Discussion}\label{discussion}

      Our 
      probabilistic approach to the distributed modeling of shallow
      landslides proved effective in the two study areas where it was tested
      (Figures~\ref{fig02} and \ref{fig07}). In both areas, the maps showing
      the geographical distribution of the factor of safety
      $F_{\mathrm{S}}$ obtained using TRIGRS-P were better predictors of the
      distributions of known rainfall-induced landslides than the
      corresponding maps obtained adopting the original TRIGRS
      approach. This conclusion is supported by the indices used to measure
      the forecasting skills of the different models, and particularly the
      area under the ROC, AUC (Tables~\ref{tab02}--\ref{tab04} for
      Mukilteo, and Table~\ref{tab06}, \ref{tab07}, \ref{tab08} for
      Frontignano). The runs in which we allowed a~large variability of the
      input parameters (e.g. $\lambda =0.50$ or $\lambda =0.75$) were better
      predictors of the geographical distribution of known landslides than
      the models prepared using a~reduced variability in the model
      parameters (e.g. $\lambda =0.1$) \citep{guzzetti2006a,rossi2010}.
      This is shown in the insets in Figure~\ref{fig04}, where a~portion of
      the results for the Mukilteo study area is shown at a~larger
      scale. The variability of the geographical distribution of the
      $F_{\mathrm{S}}$ is also shown in Figure~\ref{fig12} where we have
      plotted the minimum, the maximum, and the standard deviation of the
      computed $F_{\mathrm{S}}$ values. In particular, the map of the
      standard deviation provides quantitative and spatially distributed
      evidence of the uncertainty associated with the distributed modeling
      of landslide instability.

      We studied the variation of the computed factor of
      safety. Figure~\ref{fig13} shows histograms for the distribution of
      the values of the factor of safety $F_{\mathrm{S}}$ in selected grid
      cells in the Mukilteo (Figure~\ref{fig13}a--c) and the Frontignano
      (Figure~\ref{fig13}d--f) study areas. For simplicity, in the Figure we
      show the results obtained for a~single lithological type i.e., the
      transition sediments (Qtb, indicated as unit $1$ in
      Table~\ref{tab01}) in the Mukilteo area (Figure~\ref{fig03}a), and the
      sand-silt-clay (unit~5 in Table~\ref{tab05}) in the Frontignano area
      (Figure~\ref{fig08}a). Results for other lithological types in the two
      study areas are similar. We adopted the following procedure to obtain
      the histograms. First, we performed $100$ probabilistic simulations to
      obtain a~large set of values of the factor of safety $F_{\mathrm{S}}$,
      and we computed the average value of the factor of safety,
      $\overline{F_{\mathrm{S}}}$ for each grid cell in the two modeling
      domains. For both study areas, a~value of $\lambda = 0.50$ (and $\nu
      =1.0$) was used for the variability of the geotechnical and hydrological 
      parameters. Next, we selected three subsets of 1000~grid cells, with
      $0<\overline{F}_{\mathrm{S}} \le 1.5$, $1.5< \overline{F}_{\mathrm{S}}
      \le 3.0$, and $\overline{F}_{\mathrm{S}}>3$, respectively. Finally, we
      used all the computed values of the $F_{\mathrm{S}}$ in each subset to
      construct the histograms. Inspection of the histograms reveals that
      for $\overline{F}_{\mathrm{S}}>3$ (Figure~\ref{fig13}c,~f) the
      distribution of the predicted factor of safety is almost uniform and 
      does not show a~predominant value. 
      Instead, for
      $\overline{F}_{\mathrm{S}}<1.5$ the distribution of the predicted
      factors of safety peaks at $F_{\mathrm{S}} \approx 1.0$
      (Figure~\ref{fig13}a,~d). For $1.5<\overline{F}_{\mathrm{S}}\le 3.0$,
      results are intermediate (Figure~\ref{fig13}b,~e).

      In conclusion, the probabilistic approach results in a~number of model 
      outputs, each representing the geographical distribution of the $F_{\mathrm{S}}$ 
      values. In this work, 16~runs were performed. Availability
      of multiple results allows for the analysis of the sensitivity of the
      \begin{table*}[!htp]
        \caption{Geotechnical parameters for the geological units cropping out
          in the Frontignano study area (Figure~\ref{fig08}a). $c$, cohesion; $\phi$,
          friction angle; $D_{0}$, hydraulic diffusivity; $K_{\text{S}}$, saturated hydraulic
          conductivity; $\theta_{\mathrm{s}}$, saturated water content; $\theta_{\text{r}}$,
          residual water content; $\alpha$, inverse of capillary fringe.
          Geological units: $1$, sand; $2$, clay; $3$, flysch deposits; $4$,
          gravel, sand, silt, and clay; $5$, sand, silt, and clay.}
                {
                  \begin{tabular}{lccccccc}
                    \tophline
                    Unit & $c$ & $\phi$ & $D_0$ & $K_{\mathrm{s}}$ & $\theta_{\mathrm{s}}$ & $\theta_{\mathrm{r}}$ & $\alpha$ \\
                    & [kPa] & [deg] & [m$^2$\,s$^{-1}$] & [m\,s$^{-1}$] & -- & -- & [m$^{-1}$] \\
                    \middlehline
                    $1$ & $3.0$ & $31$ & $3.8 \cdot 10^{-4}$ & $1.0 \cdot 10^{-4}$ & $0.20$ & $0.05$ & $2$ \\
                    $2$ & $4.0$ & $18$ & $5.0 \cdot 10^{-6}$ & $1.0 \cdot 10^{-7}$ & $0.80$ & $0.07$ & $5$ \\
                    $3$ & $50.0$ & $25$ & $8.3 \cdot 10^{-6}$ & $1.0 \cdot 10^{-6}$ & $0.45$ & $0.1$ & $5$ \\
                    $4$ & $15.0$ & $30$ & $4.0 \cdot 10^{-4}$ & $1.0 \cdot 10^{-4}$ & $0.45$ & $0.1$ & $5$ \\
                    $5$ & $3.0$ & $15$ & $4.7 \cdot 10^{-3}$ & $1.0 \cdot 10^{-4}$ & $0.50$ & $0.1$ & $1$ \\
                    \bottomhline
                \end{tabular}}
                \belowtable{%
                }\label{tab05}
      \end{table*}
      model to variations in the input parameters controlling the stability
      conditions. Variability depends on multiple causes \citep{uchida2011},
      including: (i) the
      natural variability in the geotechnical and hydrological properties of
      the soils; (ii) the inability of determining accurate values for the
      geotechnical and hydrological parameters, and (iii) the fact that the
      models are simplified and do not represent the natural (physical)
      conditions in the study area.

      The probabilistic approach allowed the investigation of the combined
      effects of the natural variability inherent in the model parameters,
      and of the uncertainty associated with their definition over large
      areas. However, the approach cannot separate the two causes for the
      variability. Also, the probabilistic approach cannot validate the
      physics in the model better than the deterministic approach. It should
      be noted that in our runs with probabilistic input parameters, the
      geotechnical and hydrological properties were treated explicitly as
      independent (uncorrelated) variables. This was a~simplification. In
      reality, some dependence (correlation) exists between the different
      geo-hydrological properties. As an example, the saturated water
      content ${\theta }_{\mathrm{s}}$ affects the saturated hydraulic
      conductivity $K_{\mathrm{s}}$ and the hydraulic diffusivity
      $D_0$. However, selection of values for the different properties based
      on field tests, laboratory experiments, or through a~literature search
      resulted in values for the considered properties that were implicitly
      dependent. This is because e.g., cohesion, angle of internal friction,
      soil unit weight, and hydraulic conductivity depend one upon the
      other. Furthermore, no spatial correlation of the individual variables
      was considered in the modeling. This was also a~simplification,
      because spatial correlation exists between the geo-hydrological
      properties \citep[e.g.,][]{rodrigueziturbe1999,western2004}. Adoption
      of the uniform distribution to determine the possible range of
      \begin{table}[!hbp]
        \caption{As in Table~\ref{tab02}, but for the Frontignano area.}
                {
                  \begin{tabular}{lcccc}
                    \tophline
                    Model type & TPR & FPR & ACC & PPV \\
                    \middlehline
                    Saturated & $0.42$ & $0.25$ & $0.75$ & $0.02$ \\
                    Unsaturated & $0.18$ & $0.13$ & $0.86$ & $0.02$ \\
                    \bottomhline
                \end{tabular}}
                \belowtable{%
                }\label{tab06}
      \end{table}
      variation of the individual parameters, combined with the accepted
      modeling simplifications, has resulted in more ``extreme'' results,
      but not in unrealistic results.

      Results of our approach were obtained adopting the uniform
      distribution to describe the uncertainty associated with the
      geo-hydrological parameters. TRIGRS-P allows for the use of the
      Gaussian and the uniform distributions. In the runs presented in this
      work, we explored only part of the variability associated with the
      physical model describing slope instability forced by rainfall
      infiltration (Figure~\ref{fig01}b), and specifically the variability
      associated with the mechanical and hydrological parameters of the
      materials involved in the hypothetical landslides. We did not consider
      the local morphological variability, e.g.  the uncertainty in the
      description of the terrain given by the DEMs. Terrain gradient is an
      important parameter for the computation of the factor of safety
      $F_{\mathrm{S}}$. Inspection of Equation~(\ref{eq01}) shows that
      variability in the terrain gradient $\delta $ results in variability
      in the local stability conditions, measured by
      $F_{\mathrm{S}}$. Furthermore, in our runs soil depth was
      a~(non-linear) function of the local slope
      \citep{derose1996,salciarini2006}. Variations in the slope will result
      in variations in soil thickness, and in the local stability
      conditions. Preliminary results obtained adding a~uniform random
      perturbation to the DEM for the Frontignano area confirmed the (large)
      sensitivity of the physically based models to the topographic
      information \citep{montgomery1994,vanwesten2008,tarolli2012}.

      Rainfall history and geographical pattern also control the local
      stability/instability conditions, and their temporal and spatial variations. 
      For Mukilteo, we used the measured rainfall history that triggered shallow
      \begin{table}[!hbp]
        \caption{As in Table~\ref{tab03}, but for the Frontignano area.}
                {
                  \begin{tabular}{lccccc}
                    \tophline
                    $\lambda $ & TPR & FPR & ACC & PPV & AUC\\
                    \middlehline
                    $0.01$ & $0.42$ & $0.25$ & $0.75$ & $0.03$ & $0.59$ \\
                    $0.10$ & $0.41$ & $0.25$ & $0.75$ & $0.03$ & $0.60$ \\
                    $0.50$ & $0.37$ & $0.22$ & $0.77$ & $0.03$ & $0.64$ \\
                    $0.75$ & $0.27$ & $0.17$ & $0.83$ & $0.03$ & $0.64$ \\
                    $1.00$ & $0.05$ & $0.04$ & $0.95$ & $0.02$ & $0.63$ \\
                    \bottomhline
                \end{tabular}}
                \belowtable{%
                }  \label{tab07}
      \end{table}
      \begin{table}[!htp]
        \caption{As in Table~\ref{tab04}, but for the Frontignano area.}
                {
                  \begin{tabular}{lccccc}
                    \tophline
                    $\nu $ & TPR & FPR & ACC & PPV & AUC\\
                    \middlehline
                    $0.8$ & $0.57$ & $0.37$ & $0.63$ & $0.02$ & $0.65$ \\
                    $0.9$ & $0.44$ & $0.27$ & $0.72$ & $0.02$ & $0.64$ \\
                    $1.0$ & $0.27$ & $0.17$ & $0.83$ & $0.03$ & $0.64$ \\
                    $1.1$ & $0.10$ & $0.06$ & $0.92$ & $0.02$ & $0.64$ \\
                    \bottomhline
                \end{tabular}}
                \belowtable{%
                } \label{tab08}
      \end{table}
      landslides in the winter 1996--1997. For Frontignano, we used the
      rainfall history that has resulted in shallow landslides in the winter
      2004--2005. However, sensitivity of the models to the temporal and
      spatial variation of rainfall was not investigated, as this was 
      not within the scope of the work. 
      The effects of changing 
      rainfall histories was investigated by \citet{alvioli2014}, who examined 
      storms of different durations and average intensities.
      The rainfall data used in the two runs were obtained from rain gauges located in the
      vicinity of the study areas. The rainfall measurements may not
      represent the exact amount of rainfall at each grid cell in the
      modeling domain. We further assumed a~uniformly distributed rainfall
      in the geographical modeling domains. Runs performed in the
      Frontignano area adopting different rainfall histories (e.g., (i)
      a~uniform rainfall rate of $0.36\,\unit{mm\,h^{-1}}$ for a~$4$-week period,
      for a~cumulated rainfall $E= 242\,\unit{mm}$, (ii) a~single rainfall event
      with $5\,\unit{mm\,h^{-1}}$ for 24\,\unit{h}, $E = 121\,\unit{mm}$, and (iii)
      intermittent $3$-day rainfall periods with $I = 1.0\,\unit{mm\,h^{-1}}$
      separated by $4$-day dry periods, for a~$4$-week period, $E = 288\,\unit{mm}$) 
      revealed that the geographical distributions of the
      $F_{\mathrm{S}}$ obtained with the different rainfall histories were
      similar. However, the local instability conditions ($F_{\mathrm{S}}\le
      1$) were reached at different times. The difference may be significant
      if the model results are used in a~landslide early warning system
      \citep{aleotti2004,godt2006}. We did not evaluate the sensitivity of
      the model parameters to the different rainfall histories.

      It should be noted that the probabilistic approach of 
      TRIGRS-P could be used to infer reasonable values of the parameters
      describing terrain characteristics, where they are largely unknown, by
      exploring a~large parameter space in a~random way and comparing with
      known distributions of landslides.

      Adoption of a~probabilistic approach with multiple runs using randomly
      generated different set of input parameters results in longer computer
      processing times. The time required for a~single TRIGRS-P simulation
      is only slightly longer than the time needed for the corresponding
      TRIGRS simulation, since the random variables were computed before
      running the slope stability and infiltration model. The time for this
      initial step depends on the size (in grid cells) and complexity of the
      modeling domain. The processing time of the multiple runs required by
      the TRIGRS-P approach to have a~statistical significance may be easily
      reduced by exploiting the multi-core architecture of modern CPUs, just
      running simultaneously multiple instances of the TRIGRS-P code
      initialised with different sets of parameters.  Since our aim is to
      eventually use the TRIGRS as a~region-wide and possibly nation-wide
      early warning system, we give an estimate of the computing resources
      required. Using the same spatial resolution, a~larger area will
      require a~larger processing time, with the time increasing linearly
      with the number of grid cells. The time required for a~simulation
      depends also on rainfall history. A~more complex history
      (i.e., a~shorter step between two subsequent inputs of rainfall
      intensity) will result in a~longer processing time, with time
      increasing with the square of the time steps. Finally, processing time
      depends on the type of hydrological model used, with the saturated
      model requiring roughly half the time of the unsaturated model.

      When using the probabilistic approach, we adopted a~strategy based on
      a~convergence level, $\eta$. First, we computed two probabilistic sets
      with $n$ and $m > n$ simulations. Next, for the two independent sets
      and for each grid cell, we computed the mean of the factor of safety
      $\overline{F}_{\mathrm{S}}$. Then, we obtained the difference of the
      mean values of the factor of safety $\Delta \overline{F}_{\mathrm{S}}$
      for each cell, and we identified the maximum value of $\max(\Delta
      \overline{F}_{\mathrm{S}})$ in the modeling domain. If $\max(\Delta
      \overline{F}_{\mathrm{S}})\le \eta $, the convergence level was
      reached and no additional simulations were performed. Instead, if
      $\max(\Delta \overline{F}_{\mathrm{S}})>\eta $ convergence was not
      reached, a~larger probabilistic set was prepared, and the test
      repeated. In our two study areas 16 simulations were sufficient to
      obtain a~convergence level $\eta =0.05$. This level was considered
      adequate for the two study areas. This may not be the case in other
      areas, in significantly large areas, or in areas characterized by
      a~larger physiographical variability. For simulations covering large
      areas, we hypothesized areas extending between $10^{1}$ and
      $10^{5}\,\unit{km^{2}}$ with grids of resolution from $1\,\unit{m} \times 1\,\unit{m}$ to
      $30\,\unit{m} \times 30\,\unit{m}$, and computed the memory usage and execution
      time for (i) a~single deterministic simulation adopting a~saturated
      soil model (Figure~\ref{fig14}a), and (ii) a~probabilistic set of 16
      simulations using an unsaturated soil model (Figure~\ref{fig14}b).

      Since the TRIGRS (and TRIGRS-P) model uses a~cell-by-cell description
      of the study area, and the equations describing the stability of each
      cell are independent from the neighbouring cells behavior, the code is
      most suited for a~parallel implementation using MPI libraries. We
      performed preliminary simulations, showing that a~significant speedup
      ($\simeq 1/N$, with N the number of processing elements used) can be
      obtained for the computing-intensive portions of the code. One problem
      associated with significantly large areas is the use of
      memory. In a~truly parallel implementation of the code, each computing
      element or core should load into memory only the portion of data
      relevant to its task, which is currently not implemented. 

\conclusions\label{conclusions}

      We prepared a~probabilistic version of the Transient Rainfall
      Infiltration and Grid-Based Regional Slope-Stability Analysis code,
      TRIGRS \citep{baum2002,baum2008}, and tested the new code TRIGRS-P in
      two study areas: Mukilteo, near Seattle, USA, and Frontignano, near
      Perugia, Italy. The tests suggest that the runs initialized with
      random values of the input parameters generated according to proper
      probability distribution functions, were better predictors of the
      spatial location of rainfall-induced shallow landslides than the
      corresponding original TRIGRS runs.  This was measured by different
      metrics used to evaluate the comparison of the spatial forecasts of
      the instability conditions ($F_{\mathrm{S}}$ values) against maps
      showing recent rainfall-induced landslides, in the two study
      areas. Adoption of a~probabilistic-initiated framework allowed the
      investigation of the sensitivity of the model used to determine the
      stability conditions to the geotechnical and hydrological properties
      of the terrains where landslides can develop. The observed sensitivity
      was attributed to the combined effect of the natural variability
      inherent to the geotechnical and hydrological properties of the slope
      materials, and to the fact that the numerical model is an approximate
      representation of the complex processes controlling rainfall-induced
      slope instability in an area. However, the probabilistic approach cannot
      separate the two causes of variability. Probabilistic modeling of
      rainfall-induced shallow landslides requires longer processing times,
      when compared to the corresponding deterministic
      modeling. A~parametric study proved that the approach is
      computationally feasible even for very large areas ($10^{4}\,\unit{km^{2}}$,
      $10^{8}$~grid cells) if a~computer grid is used, and a~parallel
      computing strategy is adopted. We expect the probabilistic approach to
      improve the current capability to forecast the occurrence of rainfall-induced 
      shallow landslides, and to facilitate the investigation of the
      variability of slope material properties over large areas.

\appendix

      \section{Formulation of the model}\label{app:appa}

      In this appendix we summarize the solutions of Equation~(\ref{eq02})
      implemented in deterministic code TRIGRS
      \citep{baum2010}. Approximations are given for: (i) unsaturated soil
      conditions \citep{srivastava1991}, (ii) saturated soil conditions
      \citep{iverson2000}, and (iii) a~two-layer soil model \citep{baum2010}
      represented schematically in
      Figure~\ref{fig01}b. \subsection{Unsaturated soil} In their model for an
      unsaturated soil, \citet{srivastava1991} use relation (Equation~\ref{eq02})
      to linearize Equation~(\ref{eq02}). The explicit solution for the hydraulic
      conductivity (Equation~\ref{eq03}), subject to the initial and boundary
      conditions
given by Equation~(\ref{eq04}), is the following:
\begin{widetext}
  \begin{align}\label{eqA1}
    K(Z,t)=\,& I_{\text{Z}} - \left[ I_{\text{Z}}-K_{\mathrm{s}}
      \right] e^{-\alpha_1(d_{\mathrm{w}}-Z)}
    - 4 \left( I_{\text{Z}} -I_{ZLT} \right) e^{\frac{\alpha_1 Z}{2}\,-D_{\psi} \frac{t}{4}}
\,\sum_{m=1}^{\infty} \frac{\sin [\Lambda_m \alpha_1 (d_{\mathrm{w}}-Z)] \sin(\Lambda_m \alpha_1 d_{\mathrm{w}})}
          {1+\frac {\alpha_1 d_{\mathrm{w}}}{2} +2\Lambda_m^2 \alpha_1 d_{\mathrm{w}}}\,\,e^{-\Lambda_m^2 D_{\psi}t}
  \end{align}
\end{widetext}
      where ${\alpha }_{1} = \alpha\,{\cos}^{2}\delta $, $I_{ZLT}$ is
      the initial surface flux (where the subscript LT indicates the long
      term infiltration rate), $I_Z$ is the surface flux of a~given
      intensity for the considered time interval, $D_{\psi } ={\alpha
      }_{{1}}K_{\mathrm{s}} /({\theta }_{\mathrm{s}} -{\theta }_{\mathrm{r}}
      )$ and ${\Lambda }_m$ are the positive roots of the pseudoperiodic
      characteristic equation ${\text{tan}}\left(\Lambda {\alpha
      }_{{1}}d_{\mathrm{w}}\right)+2\Lambda =0$. The pressure head $\psi
      \left(Z,t\right)$ in the unsaturated zone is obtained by inversion of
      \citet{gardner1958} equation, Equation~(\ref{eq03}):
\begin{align}
\label{eqA2}
  \psi \left( Z, t \right)= \frac{\cos \delta}{\alpha_1} \ln \left[ \frac{K(Z,t)}{K_{\mathrm{s}}}  \right] + \psi_0
\end{align}

\subsection{Saturated soil}

      For wet initial conditions, \citet{iverson2000} gave an explicit
      solution of the linearized Richards equation, for long term and for
      short term behavior. The long term represents the steady component:
\begin{align}
\label{eqA3}
  \psi(z)= (z - d_{\mathrm{w}})\left[ \cos \delta - \frac{I_{ZLT}}{K_{\mathrm{s}}} \right]
\end{align}
where $z = Z{\text{cos}}\delta $ and $d_{\mathrm{w}}$ is the depth to the water table (see Figure~\ref{fig01}b). 
The short term represents the transient component:
\begin{widetext}
  \begin{align}
&\label{eqA4}
    \psi (Z, t \le T )=(Z -d_{\mathrm{w}})\beta +
    +\frac{I_Z}{K_{\mathrm{s}}}\left[ \left( \frac{\bar{D}t}{\pi}\right)^{\frac{1}{2}} e^{-\frac{Z^2}{\bar{D}t}} \right.
      \left.  -Z \,{\text{erfc}}\,\left(\frac{Z^2}{\bar{D}t} \right)  \right] \\
&    \label{eqA5}
    \psi(Z, t > T ) = \psi (Z, t \le T ) -\frac{I_Z}{K_{\mathrm{s}}} \left[ \left( \frac{\bar{D}(t-T)}{\pi} \right) ^{\frac{1}{2}}
      e^{-\frac{Z^2}{\bar{D}(t-T)}}\right. \left. -Z \,\,{\text{erfc}}\,\, \left( \frac{Z^2}{\bar{D}(t-T)} \right)  \right]
  \end{align}
\end{widetext}
\begin{table*}[!htp]
\caption{Notation.}
{
\begin{tabular}{ll}
\tophline
Symbols& Description\\
\middlehline
$A$: &Upslope contributing area, $L^2$. \\
$c$:& Soil cohesion, $ML^{-1} T^{-2}$. \\
$C$:& Specific moisture capacity, $L^{-1}$. \\
$C_0$:& Moisture capacity at saturation, $L^{-1}$.  \\
$d_{\mathrm{w}}$:& Depth to the water table, $L$. \\
$d_{\text{fp}}$:& Depth to the sliding plane, $L$.  \\
$D_0$:& Saturated hydraulic diffusivity, $L^2T^{-1}$.  \\
$E$:& Cumulated event rainfall, $L$.  \\
$F_{\mathrm{S}}$:& Factor of safety [--]. \\
$\overline{F}_{\mathrm{S}}$:& Average value of the factor of safety [--]. \\
$I$:& Mean rainfall intensity, $LT^{-1}$. \\
$K_x$:& Hydraulic conductivity in $x$ direction, $LT^{-1}$. \\
$K_y$:& Hydraulic conductivity in $y$ direction, $LT^{-1}$. \\
$K_z$:& Hydraulic conductivity in $z$ direction, $LT^{-1}$. \\
$K_{\mathrm{s}}$:& Saturated hydraulic conductivity, $LT^{-1}$. \\
$K^{*}$:& Normalized hydraulic conductivity [--]. \\
$n$:& Number of simulations in a~probabilistic set. \\
$t$:& Time, $T$. \\
$t^{*}$:& Normalized time [--]. \\
${\mathcal N} (0,1)$:& Standard normal distribution. \\
${\mathcal U} (0,1)$:& Standard uniform distribution. \\
${\mathcal N}(\overline{x},{\sigma }_x)$:& Normal distribution. \\
${\mathcal U}(\overline{y},\lambda )$:& Uniform distribution. \\
$x$:& Slope parallel coordinate, $L$. \\
$\overline{x}$:& Mean value of the generic variable $x$. \\
$y$:& Slope parallel coordinate, orthogonal to $x$, $L$. \\
$\overline{y}$:& Mean value of the generic variable $y$, $L$. \\
$y_a,y_b$:& Minimum and maximum values for the generic variable $y$, $L$. \\
$z$:& Slope normal coordinate, $L$. \\
$z^{*}$:& Normalized slope normal coordinate [--]. \\
$Z$:& Vertical coordinate, $Z = z /\cos\delta $, $L$. \\
$\alpha $:& Parameter for fitting soil-water characteristic curve, $L^{-1}$. \\
${\gamma }_{\mathrm{w}}$:& Unit weight of water,  $ML^{-2}T^{-2}$. \\
${\gamma }_{\mathrm{s}}$:& Unit weight of soil $ML^{-2}T^{-2}$. \\
$\delta $:& Slope angle, corresponds to gradient of the sliding plane [--]. \\
$\varepsilon $:& Ratio of length scales for slope-normal and lateral infiltration. \\
$\theta $:& Volumetric water content [--]. \\
${\theta }_{\mathrm{s}}$:& Saturated water content [--]. \\
${\theta }_{\mathrm{r}}$:& Residual water content [--]. \\
$\lambda $:& Range of the random variable $y$, $L$. \\
$\sigma $:& Standard deviation for a~normally distributed variable, $L$. \\
${\sigma }_x$:& Standard deviation for the normally distributed variable $x$, $L$. \\
$\phi$:& Soil friction angle [--]. \\
$\psi $:& Groundwater pressure head, $L$. \\
${\psi }^{*}$:& Normalized groundwater pressure head [--]. \\
$\widetilde{\psi }$:& Pressure head in Gardner's model, $L$. \\
$\xi $:& Generic random variable. \\
\bottomhline
\end{tabular}}
\belowtable{%
}\label{notation}
\end{table*}
       where $T$ is the rainfall duration, $\overline{D}=4D_0 \cos^2 \delta$
       is an effective hydraulic diffusivity, and erfc is the
       complementary error function:
\begin{align}
{\text{erfc}}\,(x)=1-\,{\text{erf}}\,(x)=\frac{2}{\sqrt{\pi}} \int_x^{\infty} e^{-t^2} \; \mathrm{d}t
\end{align}

\subsection{Two-layer soil model}

      The linearized Richards equation allows for the superposition of
      solutions. \citet{baum2002} have extended the \citet{iverson2000} and
      the \citet{srivastava1991} solutions to the case of a~time-varying
      sequence of surface fluxes with variable intensity and duration. They
      also considered an unsaturated layer of depth $d$ and depth to the top
      of the capillary fringe $d_u$ (see Figure~\ref{fig01}b). Solution of
      Equation~(\ref{eqA1}) was
generalized as follows:
\begin{widetext}
  \begin{align}
    K(Z,t) &= \sum_{n=1}^{N}H(t-t_n) \left\{ I_{nZ}-\left[
    I_{nZ}-K_\mathrm{s}
      \right] e^{-\alpha_1(d-Z)}  -4\left(I_{Z} -I_{ZLT}\right)
    e^{\frac{\alpha_1 Z}{2}} e^{-D_{\psi} \frac{t-t_n}{4}}
\,\cdot\right.\nonumber\\
&\cdot\left.
\sum_{m=1}^{\infty}
    \frac{\sin \left[ \Lambda_m \alpha_1 (d-Z) \right] \sin \left(
      \Lambda_m \alpha_1 d \right) }{1+\frac {\alpha_1 d}{2} +2\Lambda_m^2 \alpha_1 d}
    e^{-\Lambda_m^2 D_{\psi}(t-t_n)} \right\}\,+\nonumber \\
   &-\,\sum_{n=1}^{N}H(t-t_{n+1}) \left\{ I_{nZ}-\left[ I_{nZ}-K_\mathrm{s}
      \right] e^{-\alpha_1(d-Z)} -4\left(I_{Z} -I_{ZLT}\right)  \cdot \nonumber\right. \\
    \label{eqA6}
   & \left.\cdot\,e^{\frac{\alpha_1 Z}{2}}e^{-D_{\psi}\frac{t-t_{n+1}}{4}} \sum_{m=1}^{\infty} \frac{\sin \left[
        \Lambda_m \alpha_1 (d-Z) \right] \sin \left( \Lambda_m \alpha_1 d \right)}
          {1+\frac {\alpha_1 d}{2} +2\Lambda_m^2 \alpha_1 d}
    \,e^{-\Lambda_m^2 D_{\psi}(t-t_{n+1})} \right\}
  \end{align}
\end{widetext}
      where $I_{nZ}$ is the surface flux of a~given intensity for the $n$-th
      time interval, and $H(t - t_n)$ is the Heaviside step function.
The \citet{iverson2000} solutions of Eqs.~(\ref{eqA4}) and (\ref{eqA5}) are generalized as:
\begin{widetext}
  \begin{align}
\label{eqA7}
    \psi(Z, t) =\,& (Z-d) \beta +2 \sum_{n=1}^{\text{N}}\frac{I_{nZ}}{K_{\mathrm{s}}} \left\{ H(t-t_n)\left[D_1(t-t_n)^{\frac{1}{2}} \right] i\,{\text{erfc}}\,\,
    \left[ \frac{Z}{2[D_1(t-t_n)]^{\frac{1}{2}}}\right] \right\}\,+ \nonumber \\
    & -2 \sum_{n=1}^{\text{N}}\frac{I_{nZ}}{K_{\mathrm{s}}} \left\{
H(t-t_{n+1})\left[D_1(t-t_{n+1})^{\frac{1}{2}} \right]\, i\,{\text{erfc}}\,\,
    \left[ \frac{Z}{2[D_1(t-t_{n+1})]^{\frac{1}{2}}} \right] \right\}
  \end{align}
\end{widetext}
with ${\text{ierfc}}(\eta )=\frac{1}{\sqrt{\pi }}\exp(-{\eta }^{{2}})-\eta
{\text{erfc}}(\eta )$.

\Supplementary{zip}

\begin{acknowledgements}
      SR and MR supported by a~grant of the Italian national Department for
      Civil Protection (DPC); MA supported by Regione Umbria under contract
      \textit{por-fesr Umbria 2007--2013, asse ii, attivit\`a~a1, azione 5}
      and DPC. We thank the ENEA-Grid team (\url{www.eneagrid.enea.it}), and
      the CRESCO computer centre (\url{www.cresco.enea.it}) for support and
      the possibility of using their Grid computing facilities. We are
      grateful to L. Brakefield and A. Frankel (USGS) for their constructive
      reviews on an earlier version of this manuscript.
\end{acknowledgements}

\newpage

\end{document}